\documentclass[reprint, nofootinbib, amsmath,amssymb, aps, floatfix, 
twoside]{revtex4-1}

\usepackage[utf8]{inputenc}
\usepackage[english]{babel}
\usepackage[T1]{fontenc}

\usepackage{array}
\usepackage{amsmath}
\usepackage{amsfonts}
\usepackage{amssymb}
\usepackage{amstext}
\usepackage{amsthm}

\usepackage{color}
\usepackage{graphicx}
\usepackage[colorlinks=true,citecolor=blue,linkcolor=blue,urlcolor=blue]{hyperref}
\usepackage{mathtools}
\usepackage{times}
\usepackage{latexsym}
\usepackage[running,mathlines]{lineno}
\usepackage{verbatim}
\usepackage{physics}
\usepackage{bm}
\usepackage{ulem}
\usepackage[most]{tcolorbox}

\usepackage{listings}
\usepackage{xcolor}  
\usepackage{ragged2e}

\definecolor{codegreen}{rgb}{0,0.6,0}
\definecolor{codegray}{rgb}{0.5,0.5,0.5}
\definecolor{codepurple}{rgb}{0.58,0,0.82}
\definecolor{backcolour}{rgb}{0.95,0.95,0.92}

\lstdefinestyle{mystyle}{
    backgroundcolor=\color{backcolour},   
    commentstyle=\color{codegreen},
    keywordstyle=\color{magenta},
    numberstyle=\tiny\color{codegray},
    stringstyle=\color{codepurple},
    basicstyle=\ttfamily\footnotesize,
    breakatwhitespace=false,         
    breaklines=true,                 
    captionpos=b,                    
    keepspaces=true,                 
    numbers=left,                    
    numbersep=5pt,                  
    showspaces=false,                
    showstringspaces=false,
    showtabs=false,                  
    tabsize=2
}

\newtheoremstyle{mytheoremstyle} 
  {0.5cm}                        
  {0.5cm}                        
  {}                     
  {}                             
  {\bfseries}                    
  {.}                            
  {0.5em}                        
  {}                             
\theoremstyle{mytheoremstyle}
\newtheorem{theorem}{Theorem}[section] 
\theoremstyle{mytheoremstyle}
\theoremstyle{mytheoremstyle}

\begin{document}
\raggedbottom

\title{\Large{An introduction to Neural Networks for Physicists}}

\author{Gustavo Café de Miranda$^\dagger $} \email[Email:]{ gcaf0125@uni.sydney.edu.au}
\affiliation{University of Sydney, Sydney, NSW, Austrália}

\author{Gubio G. de Lima$^\dagger $}
\affiliation{Universidade Federal de São Carlos, Departamento de Física, São Carlos, SP, Brasil}

\author{Tiago de S. Farias }
\affiliation{Universidade Federal de São Carlos, Departamento de Física, São Carlos, SP, Brasil}

 \address{$^\dagger $ Both Authors contributed equally to the present work.}

\bigbreak 

\begin{abstract}
    
Machine learning techniques have emerged as powerful tools to tackle various challenges. The integration of machine learning methods with Physics has led to innovative approaches in understanding, controlling, and simulating physical phenomena. This article aims to provide a practical introduction to neural network and their basic concepts. It presents some perspectives on recent advances at the intersection of machine learning models with physical systems. We introduce practical material to guide the reader in taking their first steps in applying neural network to Physics problems. As an illustrative example, we provide four applications of increasing complexity for the problem of a simple pendulum, namely: parameter fitting of the pendulum's ODE for the small-angle approximation; Application of Physics-Inspired Neural Networks (PINNs) to find solutions of the pendulum's ODE in the small-angle regime; Autoencoders applied to an image dataset of the pendulum's oscillations for estimating the dimensionality of the parameter space in this physical system; and the use of Sparse Identification of Non-Linear Dynamics (SINDy) architectures for model discovery and analytical expressions for the nonlinear pendulum problem (large angles).

\begin{description}
\item[Keywords] Classical Physics; Neural networks; Tutorial.
\end{description}
\end{abstract}

\maketitle
\section{Introduction}

Machine Learning (ML) has gained considerable attention in recent years due to its success in commercial, industrial, and especially service-sector applications \cite[p.27, p.vi]{geron, NathanKutz2017}. Today, intelligent algorithms—often associated with the broader term Artificial Intelligence (AI)—are deeply integrated into digital technologies. These techniques have proven highly effective in processing large volumes of data, facilitating the solution of complex problems through statistical methods \cite[p.2]{carleo_machine_2019}. This success has attracted the attention of researchers, establishing ML as a powerful statistical tool and a potential ally in scientific discovery.

Although a promising resource, ML often produces models that are difficult to interpret, commonly referred to as ``black boxes'' because the causal relationships that leads the algorithm to their predictions cannot be clearly determined. This opacity has led to skepticism within the scientific community, who tend to favor more interpretable models with clear theoretical foundations \cite[p.2]{carleo_machine_2019}. However, such skepticism has not hindered research into novel ML algorithms, particularly in fields that prioritize applications. As a result, ML has established itself as a promising tool for advancing scientific inquiry. The following paragraphs outline some of the key areas where ML has proven valuable in the scientific domain, with a particular emphasis on applications in Physics.

Although ML has gained significant attention in recent years, many of its foundational models were developed and refined over several decades \cite{Schmidhuber2015}. From the second half of the 20th century onward, improvements in computational power, statistical techniques, and mathematical modeling laid the groundwork for modern ML. Among the various approaches that emerged, one particularly influential development was the creation of artificial Neural Networks (NN), inspired by studies on biological neurons. These algorithms were designed to replicate, in a simplified way, the learning processes observed in living organisms.

Artificial neural networks rely on different training paradigms that determine how models interact with data to solve specific problems. These paradigms—supervised learning, reinforcement learning, and unsupervised learning—each define a distinct approach to extracting meaningful patterns from data and have been applied in a wide range of fields.

In supervised learning, models are trained on labeled datasets, where each input is associated with a known output. This allows the algorithm to approximate a function that maps inputs to outputs \cite[p.30]{geron}. Some of the most widely used supervised learning algorithms include\footnote{In English: K-Nearest Neighbors, Logistic Regression, Support Vector Machines (SVM), Decision Trees, and Artificial Neural Networks.}: K-Nearest Neighbors \cite{KNN,KNN2}, Logistic Regression \cite{logiste-regression1,logiste-regression2}, Support Vector Machines \cite{SVM1,SVM2}, Decision Trees \cite{decision-tree1,decision-tree2}, and Artificial Neural Networks \cite[p.29-47]{supervised-nn}.

In contrast, unsupervised learning deals with algorithms that analyze unlabeled data, meaning that the underlying structure or patterns in the data are not explicitly known and must be discovered by the learning process. In this setting, models must identify relationships and structures purely from the input data, optimizing an objective function that captures intrinsic patterns. Optimizing this function yields outputs that satisfy predefined constraints, which may be either defined by the problem at hand or by training strategies. As a historical remark, unsupervised learning gained prominence following studies on neural networks inspired by biophysical models of the feline visual cortex \cite[p.10]{Schmidhuber2015}. Some widely used unsupervised learning algorithms include K-Means \cite[p.183-189]{unsupervised}, DBSCAN \cite[p.199]{unsupervised}, PCA \cite[p.56]{unsupervised}, t-Distributed Stochastic Neighbor Embedding (t-SNE) \cite[p.13-15]{supervised-nn}, and autoencoders \cite{Marquardt2021,Goodfellow-et-al-2016,SINDyAutoencoder_steven2019}.

Reinforcement learning takes a different approach, where a model—typically called an agent—learns to make decisions by interacting with a dynamic environment through trial and error. The agent observes the state of the environment, performs actions, and receives rewards or penalties based on the effectiveness of its choices. The objective is to develop an optimal strategy that maximizes cumulative rewards over time. Reinforcement learning has been successfully applied in diverse domains, including strategic game-playing, robotics, autonomous control systems, and recommendation engines. Some of the most notable applications include AI systems that have surpassed human performance in board games and video games, such as AlphaGo \cite{alphago} and Atari \cite{atari}, as well as robotic systems that learn to navigate and manipulate objects \cite{robothand}.

\begin{figure}[ht]
    \justifying 
    \includegraphics[width=0.98\linewidth]{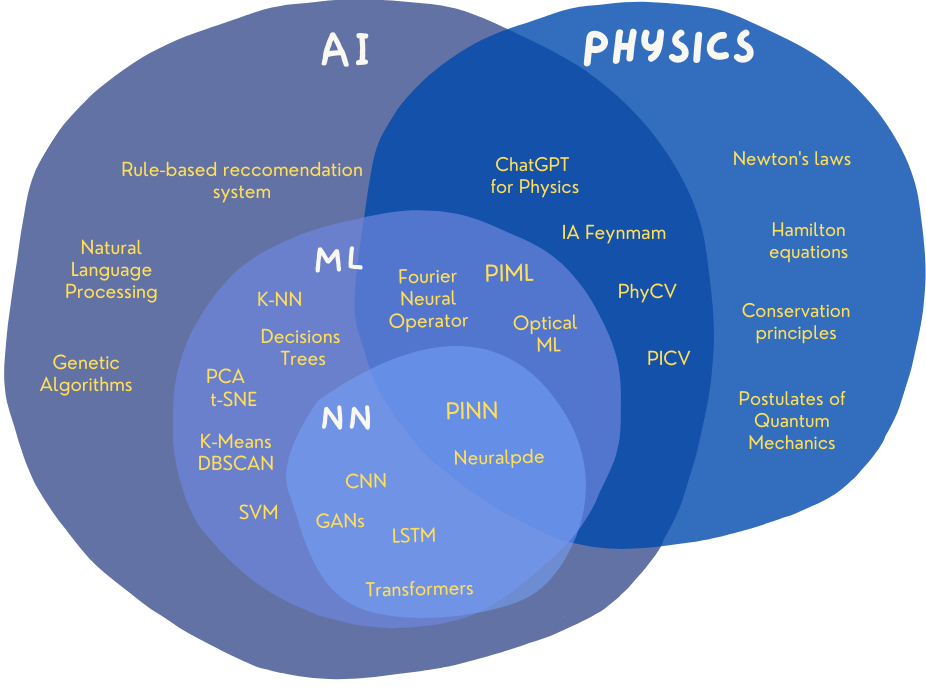}
    \caption{ \justifying Diagram illustrating the intersection of Artificial Intelligence (AI), Machine Learning (ML), Neural Networks (NN), and Physics. Various hybrid approaches emerge at these intersections, including frameworks like PINN and PIML, which integrate ML/NN with physical laws, as well as applications such as AI Feynman and PhyCV, which incorporate physics into AI algorithms to improve model accuracy. }
    \label{fig:fisica-IA}
\end{figure}

In fundamental science, Particle Physics was one of the first to adopt ML techniques as a tool, driven by the need to analyze massive real-time data streams. Early applications of ML in this field focused on managing large datasets that required rapid processing. In the field of High-energy Physics, for example, decision trees were widely used in particle collider experiments, where fast classification methods were essential for event detection and data storage \cite[p.1]{Gligorov2013}. Later, NN demonstrated superior performance in these tasks, outperforming traditional data analysis methods \cite[p.8]{baldi2014}.

Another major application of ML in Particle Physics involves theoretical models that rely on perturbative expansions in Quantum Field Theory, particle-detector interactions, and phenomenological models that describe interactions at various scales. These models often contain free parameters that must be fitted using experimental data. NN have been widely employed for parameter estimation, bridging the gap between theoretical predictions and experimental observations in both Particle Physics and Cosmology \cite[p.11]{carleo_machine_2019}\cite{Feickert}.

NN have proven effective in many computationally intensive tasks, despite the high computational cost of training. However, once trained, these models can be applied to new datasets with significantly reduced computational demands. ML has also yielded promising results in many-body physics, both classical and quantum, where phase characterization in spin models (e.g., 2D Ising models) is a key challenge \cite{carrasquila2017, VonNieuwenburg2017}. In Condensed Matter Physics and Materials Science, ML has been successfully integrated with existing methods to accelerate energy calculations in Density Functional Theory (DFT), leading to significantly faster computations compared to conventional approaches \cite{Smith2017}.

Another particularly important direction in ML research has been the integration of physical constraints into ML models to improve interpretability. This approach, known as Physics-Informed Machine Learning (PIML) \cite{PIML-review,PIML-review-nature,PIML-review2} and Physics-Informed Neural Networks (PINN) \cite{PINN-review,PINN-review2}, has been applied across various areas of Physics, including Fluid Mechanics \cite{PINN-review-fluid,PINN-review-fluid2}, uncertainty quantification \cite{Psaros}, Dynamical Systems \cite{WangandR}, Quantum Many-body Systems \cite{Carrasquilla}, Photonics \cite{nn-phonics}, and Classical Optics \cite{nn-optics}. Additionally, hybrid modeling frameworks that combine ML with physical knowledge have also been explored \cite{Pinn-hibrido0,Pinn-hibrido1,Pinn-hibrido2}. On the other hand, some studies have pursued the inverse approach: using ML to uncover physical principles. Examples include the use of symbolic regression to extract governing equations \cite{MaxTegmark2020} and advanced autodifferentiation techniques in ML frameworks \cite{Baydin-automaticdiff}.

The intersection of Physics, AI, and ML has evolved into a distinct research field of its own, represented in Figure \ref{fig:fisica-IA} where the  category of NN is also represented, highlighting some of the algorithms that emerge at the interface of these domains. The large circle on the left represents AI-related algorithms and approaches, which include natural language processing, rule-based recommendation systems, and genetic algorithms. NN encompass models such as Convolutional Neural Networks (CNNs), Generative Adversarial Networks (GANs), and Transformers. The Physics set, on the right, includes fundamental concepts such as Newton’s laws, conservation principles, Hamiltonian equations, and the postulates of quantum mechanics. The intersection between ML, NN, and Physics includes frameworks like PIML and PINN, as well as specific algorithms such as Fourier Neural Operator \cite{Fouriernn}, Optical ML \cite{OML,OML2}, and NeuralPDE \cite{NeuralPDE}. The diagram also highlights applications of AI in modeling physical systems, including ``AI Feynman'' \cite{IAfeymann}, PhyCV (\textit{Physics-inspired Computer Vision}) \cite{PhyCV}, and PICV (\textit{Physics-informed Computer Vision}) \cite{PICV}. These techniques integrate computational methods with physical principles and have also been applied for educational and computational assistance in physics (\textit{ChatGPT for Physics}).

The scientific literature at the intersection of Physics and ML is often fragmented into specialized subfields that require advanced knowledge, frequently at the graduate level, to be read and understood. To make this knowledge more accessible, the authors of this article have developed materials aimed at a broader audience, including undergraduate physics students who have completed their core coursework, as well as Engineering students. To this end, we will explore classical mechanics problems combined with NN techniques, providing an intuitive pathway to understanding these complex topics.

In this study, we discuss the fundamental concepts of NNs, starting with the Perceptron in Chapter \ref{sec:perceptron}, where we provide a step-by-step guide on how to construct and apply it. We then introduce deep neural networks in Chapter \ref{sec:RNProfunda}, expanding on previously seen key ideas and presenting the most common NN architectures. Chapter \ref{sec:Capitulo3} presents four distinct applications of neural networks using a one-dimensional pendulum model. In the first case (\ref{sec:EX:Pendulo}), we apply supervised learning to train a NN to estimate a physical constant of the system (gravitational acceleration). Next, we explore the use of deep neural networks and autodifferentiation to solve the differential equation governing the pendulum (\ref{sec:EX-EDO}). We then investigate how \textit{autoencoders} (\ref{sec:EX:AutoencoderPendulo}) can be used to infer the system’s latent space, estimate its dimensionality, and filter noise from data and images. Finally, in Section (\ref{sec:EX:SINDY}), we employ SINDy autoencoders to discover the differential equation (rather than solving it) for the nonlinear pendulum system at large angles.

\section{Fundamentals of Neural Networks}
\label{sec:capitulo2}

Historically, artificial NN were studied with the goal of modeling the behavior of biological neurons. The connection between biological phenomena and intelligence motivated research beyond neurobiology, leading to the earliest studies of neural networks as AI algorithms in 1943 \cite[p.9]{Schmidhuber2015}, although the specific terminology emerged a few years later. Between the 1960s and 2000s, new NN architectures, learning methods, and applications were proposed, many of which proved to be successful in the field of Biology \cite[p.9-23]{Schmidhuber2015}.

The 2010s marked a particularly transformative period for ML and NNs. One of the decisive factors in this progress was the widespread adoption of Graphics Processing Units (GPUs), which are designed to execute thousands of mathematical operations in parallel. Unlike traditional Central Processing Units (CPUs), which feature a limited number of cores optimized for sequential tasks, GPUs consist of hundreds or even thousands of specialized cores capable of processing large volumes of data simultaneously. This massive parallelism revolutionized the training of NN, significantly reducing the time required to adjust model parameters. Consequently, it became clear that the potential of NN had been vastly underestimated \cite[p.23,24]{Schmidhuber2015}. From that point onward, NN began outperforming traditional deterministic algorithms in various competitions and benchmarks, from image processing tasks—such as the ImageNet dataset challenge—to widely publicized events, including AlphaGo\footnote{deepmind.google/technologies/alphago/}, AlphaZero, and ChatGPT \cite{Ray2023ChatGPTAC}.
\subsection{Perceptron}
\label{sec:perceptron}

\begin{figure}[h]
    \centering
    \includegraphics[width=1.0\linewidth]{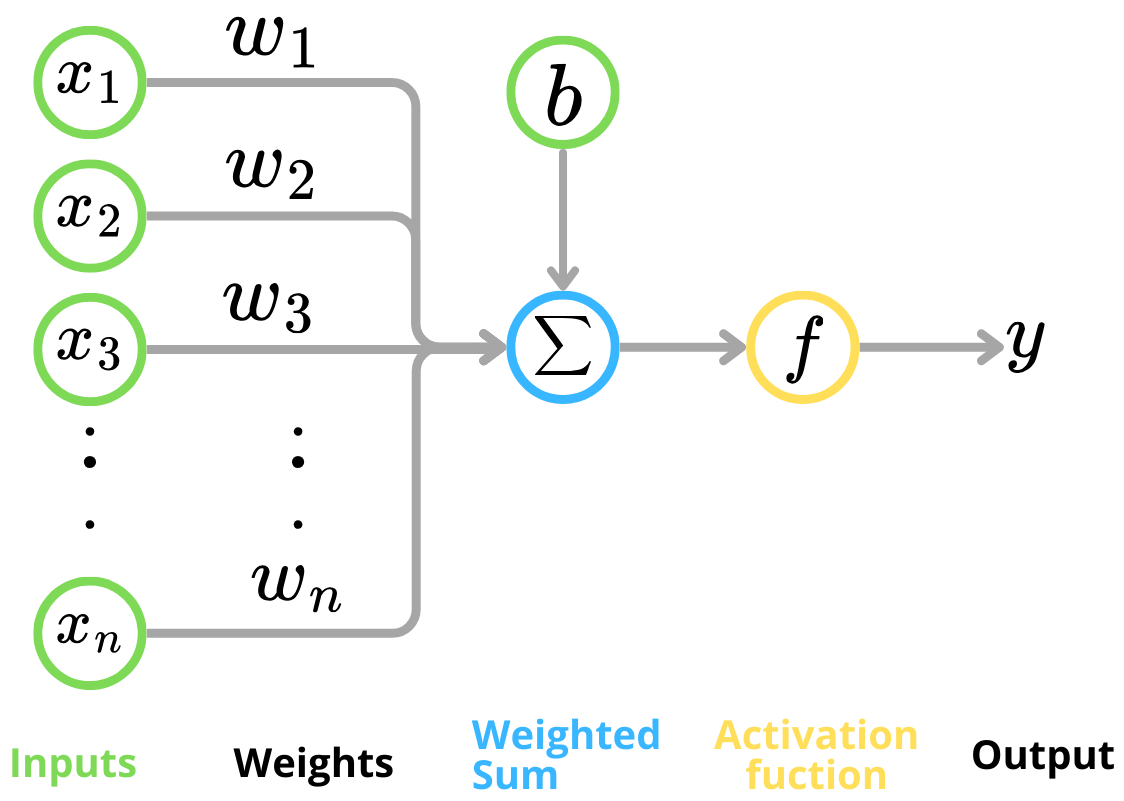}
    \caption{Illustration of the perceptron’s computational steps. The inputs $x_i$ are weighted by $w_i$, summed with a bias $b$, and then passed through an activation function $f$, producing an output $y$.} 
    \label{fig:perceptron}
\end{figure}

In artificial NN, the neuron is the fundamental unit responsible for processing information. Inspired by biological neurons, artificial neurons receive a set of inputs, apply mathematical operations, and produce an output. One of the simplest models of artificial neurons is the perceptron \cite[p.4]{Marquardt2021}. NNs are typically represented as directed graphs, where neurons correspond to vertices, and the connections between them—called synapses in the biological context and weights in the algorithmic context—are represented as edges. The perceptron can be represented as shown in Figure \ref{fig:perceptron}. 

In a simplified biological analogy, the inputs $x_i$ can be interpreted as changes in the electrostatic potential difference in the vicinity of the neuron. The weights $w_i$ represent the intensity or sensitivity of each 
connection $i \in \{1,\ldots,n\}$ with its own neighborhood. The bias $b$ can be understood either as a mean field influencing the effective field within the neuron or as an activation threshold that the neuron must exceed to become active. These elements are all taken into account by the summation operation $\Sigma$, whose output serves as the argument for the activation function $f$. This function determines whether the neuron is activated (i.e., ``fires'' a signal to connected neurons) or remains inactive. The biological analogy serves only as an illustration; in subsequent sections, we will focus exclusively on the mathematical and computational aspects of NNs. In this computational framework, the perceptron’s output $y$ represents the artificial neuron’s response to a given set of inputs, indicating whether the stimulus was sufficient to activate the neuron. 

To facilitate understanding, we will now examine each step of the perceptron’s operation in detail, covering both the mathematical formalism and its implementation in \textit{Python}.
\subsubsection{Illustrative Example: Binary Classification}

\begin{figure}[!ht]
    \centering
    \includegraphics[width=1.01\linewidth]{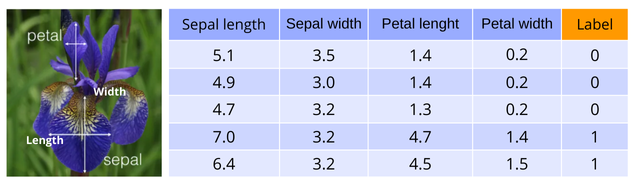}
    \caption{A sample from the Iris dataset, showing the first five rows.}
    \label{fig:iris}
\end{figure}

In this example, to illustrate the perceptron’s operation, we will analyze each component in Figure \ref{fig:perceptron}, following the order of execution of operations from left to right. These components are: inputs (data), weights, the weighted sum with the activation function, and finally, the network’s output.

\textbf{Inputs:} The first stage involves processing the input data, represented by the green circles on the left of Figure \ref{fig:perceptron}. Each element $x_i$ represents an independent variable. These inputs are numerical values characterizing the problem under investigation, e.g., we could have a dataset listing products and their features, or a database containing information about animals, plants, or people. Inputs may also originate from images \cite{imagem-nn}, audio signals \cite{sound-nn}, or text \cite{text-nn}. Additionally, the inputs to a neuron can be outputs from other neurons.

As a concrete example, we use a dataset cataloging flowers based on sepal and petal dimensions. This dataset, known as the \textit{Iris dataset} \cite{scikit-learn_iris}, is widely used in classification problems. Figure \ref{fig:iris} displays the first five rows. Each row represents a single flower, and each column corresponds to a specific feature. Mathematically, each row $j$ of the dataset can be written as a vector $\mathbf{x}_j$, :
\begin{equation}
    \mathbf{x}_j = [ x_{j1} , x_{j2}, x_{j3}, \ldots, x_{jn} ], 
\end{equation}

\noindent where $i \in \{1, \cdots, N\}$ indexes the columns ({\it features}), and $j \in \{1, \cdots, M\}$ indexes the rows (samples). With this terminology, we will call each row of the dataset a {\it data}. The final column, $N+1$, stores the {\it labels} for the flower species.

In the example shown in Figure \ref{fig:iris}, we have four features (columns) and one label per sample. For instance, the mathematical representation of the first row ($j=0$) corresponds to:
\begin{align*}
    \mathbf{x}_0 &= [ x_{01} , x_{02} , x_{03} ,  x_{04}],\\
    \mathbf{x}_0 &= [5.1, 3.5, 1.4, 0.2].
\end{align*}

To load the Iris dataset, we can use the \textit{scikit-learn} library \cite{scikit}, which provides a ready-to-use implementation, tailored for for experimenting with ML. To visualize the one element of the dataset (e.g., element $j=0$), we can use the following \textit{Python} script:

\begin{lstlisting}[language=Python, caption={}]
# Importing the dataset
from sklearn import datasets
dataset = datasets.load_iris()

j   = 0 # Sample index
x_j = dataset.data[j,:]  # Features
y_j = dataset.target[j]  # Label
print(x_j,y_j)
# Output:
[5.1, 3.5, 1.4, 0.2],  [0]
\end{lstlisting}

After accessing the dataset, the next step is initializing the NN parameters. These parameters implicitly store the knowledge acquired by the algorithm during training, as they are used to modify the information processed by each neuron. The adjustments made to these parameters define the mapping between the model's input and output, allowing the perceptron to be represented as a mathematical function of the form $g: (x_{j1}, \ldots, x_{jN}) \rightarrow y_j.$

\textbf{Weights and Bias:} The key parameters in NNs are the weights and bias. Weights ($w$) assign different levels of importance to each input feature, while the bias ($b$) is an additional term that offsets the neuron’s activation threshold, in a input-independent way, this feature enhances the adaptability and flexibility of the algorithm to a specific dataset. In the case of the perceptron, the \textit{bias} allows for shifting the decision boundary (which is either a line or a hyperplane in higher-dimensional spaces), thus improving the correct classification of inputs.

Finding the optimal way to initialize parameters is an active area of research in ML \cite[p.6]{carleo_machine_2019}, as it is dependent on the activation function used, the loss function, and specific characteristics of the problem at hand. In the absence of additional information about the system, weights can be randomly initialized using a normal distribution, but other strategies exist, such as initializing all weights close to zero, using a uniform random distribution, or applying the Glorot/Xavier initialization (suitable for symmetric activation functions) \cite{glorot}. As we will see later, what is commonly referred to as the \textit{training} of a NN is essentially an iterative update of the weight values $w_i$.  

The weight vector $\mathbf{w}$ and the input vector $\mathbf{x}$ can be written as follows:
$$
\mathbf{w} = 
\begin{bmatrix}
   w_0,& w_1, & \cdots, & w_n
\end{bmatrix},
\quad
\mathbf{x}_j = [ x_{j0} , x_{j1},  \cdots, x_{jn} ]. 
$$

In some representations, it is common to define $w_0 = b$ and write the input data such that $x_{j0} = 1$ for simplicity. We will adopt this notation. Thus, we can describe the weighted sum step (represented by the blue circle in Figure \ref{fig:perceptron}) using the dot product operation

\begin{equation*}
\mathbf{w}^T \mathbf{x}_j = \mathbf{x}^T_j\mathbf{w}  =
\begin{bmatrix}
    x_{j0}, & x_{j1}, & \cdots, & x_{jn}
\end{bmatrix}
\begin{bmatrix}
    w_0 \\ w_1 \\ \vdots \\ w_n
\end{bmatrix}.
\end{equation*}

\noindent The subscript $T$ denotes the transposed operation, so the resulting matrix multiplication expands to:

\begin{equation*}
\mathbf{x}^T_j\mathbf{w} =
x_{j0} w_0  + x_{j1} w_1  + \cdots + x_{jn} w_n  = \sum_{i=1}^{n} w_i x_{ji} + b.
\end{equation*}

\noindent More generally, we write:

\begin{equation}
    \mathbf{x}^T_j\mathbf{w} = \sum_{i=0}^{n} w_i x_{ji}.
\end{equation}

This expression represents a weighted sum of the input features for a given data point. As an illustrative example, let us consider $w = [1, 0, 0.5, -0.3], \quad b = 0.5$. The dot product operation is then computed as:

\begin{align*}
    \mathbf{w}^T \mathbf{x}_0 & = [1,5.1, 3.5, 1.4, 0.2]
    \begin{bmatrix}
       0.5 \\ 1 \\ 0 \\ 0.5 \\ -0.3
    \end{bmatrix},
\end{align*}

\begin{align*}
\mathbf{w}^T \mathbf{x}_0 &= 1 \cdot 0.5 +  5.1 \cdot 1  + 3.5 \cdot 0  +  1.4 \cdot 0.5 + 0.2 \cdot (-0.3),  \\
\mathbf{w}^T \mathbf{x}_0 &= 0.5 +  5.1  + 0  +  0.7 - 0.06, \\
\mathbf{w}^T \mathbf{x}_0 &= 6.24.
\end{align*}

This same dot product can be computed numerically with 

\noindent \begin{lstlisting}[language=Python, caption={}]
# Importing libraries
import numpy as np

w = [1, 0, 0.5, -0.3]
b = [0.5]
z = np.dot(w,x_0) + b
print(z)
# Script output
[6.24]
\end{lstlisting}

\textbf{Activation Function:} Once the weighted sum has been computed, a nonlinear function $f$, called the activation function, is applied to the result. Activation functions introduce nonlinearity into the NN, allowing the construction of deeper and more expressive models. Some commonly used activation functions include (see Table \ref{tab:activation_functions} in Appendix A for more details):

\begin{itemize}
    \item Sigmoid or Logistic: A smooth function, mainly used in the output layer of a binary classification model, as its output ranges within $[0,1]$.
    
    \item Hyperbolic Tangent (tanh): Similar to the sigmoid in shape, but its range is $[-1,1]$, with a center at zero.
    
    \item Rectified Linear Unit (ReLU): Currently one of the most widely used activation functions in deep NNs due to its computational simplicity and efficient gradient calculation while preserving nonlinearity. It outputs the input if positive and zero otherwise.
    
    \item Other types: LeakyReLU, Softmax, sine, cosine, among others \cite{activation1,activation2}.
\end{itemize}

The nonlinearity of the activation function is what enables these algorithms to approximate complex functions \cite[p.36]{dawid_modern_2022}. However, these functions must be differentiable to allow training using gradient-based optimization methods.

At the perceptron’s output stage, we obtain the following expression:
\begin{equation}
    \hat{y}_j = f\left( \sum_{i=1}^{n} w_i x_{ji} + b  \right)
\label{eq3}
\end{equation}

The output of the perceptron can be used to represent meaningful information. One of the simplest applications is a binary classifier, i.e., $\hat{y} \in \{0,1\}$, meaning the output corresponds to a decision between two possible classes (it can also be used for regression if the output is a continuous function). In our example, using the flower dataset, one can classify a flower species based on its characteristics. The original dataset contains three classes $\{0,1,2\}$ corresponding to \textit{setosa}, \textit{versicolor}, and \textit{virginica}. However, for simplicity, we reduce the dataset to two classes: 0 for \textit{setosa} and 1 for \textit{virginica}. Implementing equation \eqref{eq3} in \textit{Python}:

\noindent \begin{lstlisting}[language=Python, caption=]
# Considering the previously obtained value z
# z = [6.24]
def sigmoid_activation(z):
    f = 1/(1+np.exp(-z))
    return f
y = sigmoid_activation(z)
print(y)
# Script output
[0.998]
\end{lstlisting}

The choice of using the sigmoid activation function is motivated by its ability to constrain the perceptron’s output to the interval between 0 and 1. This property is particularly useful in binary classification problems, as the output can be interpreted as the probability of belonging to a particular class. For instance, an output close to 1 indicates a high probability that the input belongs to the positive class (1), whereas an output near 0 suggests a high probability that the input belongs to the negative class (0). This probabilistic interpretation simplifies decision-making, making the model more intuitive and allowing for statistical analysis techniques. In the example provided in the code, we observe that $y = 0.998$, indicating that, given the initialized weight values, the perceptron predicts that the data point belongs to class (1). However, when compared to the actual label, we see that this data point should belong to class (0). This discrepancy highlights an error between the perceptron’s predicted value (which has not yet been trained) and the actual value. How to measure this error and how to adjust the weights and bias to improve the output prediction are topics covered in the following subsections.

\textbf{Loss Function:} Also known as an \textit{error function} or \textit{cost function}, it is used to evaluate the performance of a model relative to real data, quantifying the discrepancy between the perceptron’s output and the expected values it should produce. By doing so, it provides a metric to be minimized during training so that it improves the algorithm’s performance. Later, we will see that this function can also be manipulated to guide the learning process of the perceptron/NN. 

In this example, we use the \textit{Mean Square Error} (MSE) loss function\footnote{Strictly speaking, in the illustrated example, we use only a single quadratic error term rather than the mean squared error, since the output vector has dimension one. However, the described case can be generalized to higher dimensions.}:

\begin{equation}
    \mathcal{L} = \sum_j^M ( y_j - \hat{y})^2,
    \label{eq-loss}
\end{equation}

where $y_j$ is the true label of the sample, and $\hat{y}$ is the model’s predicted value. Implementing equation \eqref{eq-loss} in \textit{Python}:

\noindent \begin{lstlisting}[language=Python, caption={}]
# y = [0.998], y_0 = [0]
# Quadratic error calculation for j=0
loss = (y_j - y)**2
print(loss)
# Script output
[0.99611167]
\end{lstlisting}

Note that, for the set of initial values we chose, the loss function yields a value greater than zero, i.e., far from the ideal outcome. The goal for this particular data point was to obtain $y_j=0$, but the perceptron’s prediction was $\hat{y} \approx 1$. Since the quadratic function MSE has a global minimum at ${\mathcal{L}(y_j,\hat{y}) = 0}$, we want all (or most) of the NN’s outputs, when evaluated by the loss function, to be as close as possible to this global minimum.

\textbf{Parameter Learning Rule:} The optimization of NN parameters is often referred to as ``learning'' because the adjustment process allows the NN to adapt and attempt to solve the problem it was designed for, analogous to how a person learns to solve a problem after studying and practicing. There are various techniques for adjusting NN parameters, such as genetic algorithms \cite{opt_genetical} and simulated annealing \cite{simu_anneling}, which seek optimal solutions through processes inspired by natural evolution and thermodynamics, respectively. However, gradient-based methods stand out due to their effectiveness and proven success in optimizing NN algorithms, making them widely used in practice \cite{otimization,otimization2}.

Gradient methods are based on the information given by the loss function, which measures how well the NN is performing at a given training problem. The loss function can be analogously visualized as a topographic map, where the initial position corresponds to a high point, like the top of a mountain, and the goal is to reach the lowest point, representing the function’s minimum. In this context, the gradient of the loss function provides the direction of the steepest descent, allowing the algorithm to ``observe'' its surroundings and infer decisions about the direction of each step. By iteratively applying a gradient method, the NN parameters are gradually adjusted to reduce the loss function value, moving toward the minimum.

\begin{figure} [ht]
    \centering
    \includegraphics[width=0.9\linewidth]{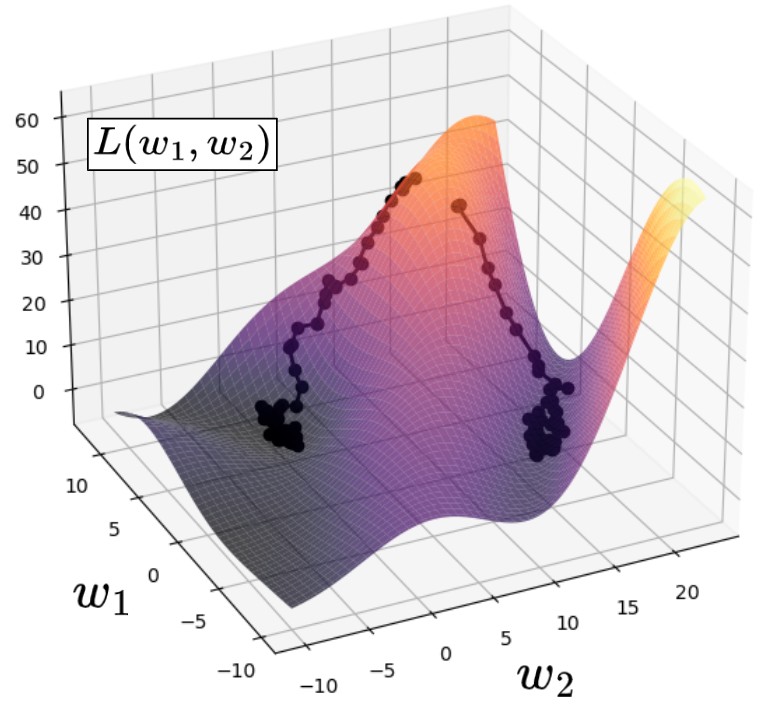}
    \caption{Graph of the parameter space $(w_1,w_2)$ (on the plane) for a loss function $L(w_1,w_2)$ (vertical axis). When minimizing a loss function, an initial parameter value is chosen, and at each iteration, the gradient descent method determines the direction to follow until the difference between iterations is smaller than a predefined tolerance. In black dots: iterations of two possible trajectories minimizing the loss function, each with different initial values, each iteration is conected to the next through a black line. Note that close initial values can lead to different minima. Image adapted from \cite[p. 80]{DissCafe}.}
    \label{fig:SGD_Topografia}
\end{figure}

One of the simplest and most well-known gradient optimization methods is gradient descent (GD). In this method, the derivative of the loss function is calculated with respect to each parameter to be updated. The parameter is then adjusted by subtracting the product of the derivative with a constant called the {\it learning rate} ($\eta$). This constant controls the step size the algorithm takes in the gradient direction. If the learning rate is too large, the algorithm may take too large steps, overshooting the desired minimum; if it is too small, the algorithm may converge too slowly.

The weight and bias updates via gradient descent are given by the following equations:
\begin{equation}
    w_{\text{new}} = w_{\text{old}} - \eta \nabla_w \mathcal{L},
    \label{eq:SGD_w_perceptron}
\end{equation}
\begin{equation}
    b_{\text{new}} = b_{\text{old}} - \eta \nabla_b \mathcal{L},
    \label{eq:SGD_b_perceptron}
\end{equation}

\noindent where $\nabla_{w,b}$ represents the gradient of the loss function with respect to the weights $w$ or bias $b$, respectively, while $\eta$ is the {\it learning rate}.

Note that the learning rate is manually defined before training begins and is generally not adjusted during the optimization process. Since it is a hyperparameter, $\eta$ requires careful selection, as an excessively high learning rate may cause the algorithm to oscillate around or even diverge from the desired minimum, while a very low rate can result in extremely slow training. The optimal learning rate can be found through cross-validation techniques or hyperparameter optimization methods, such as grid search or random search \cite{Liashchynskyi}. Choosing the correct hyperparameters is crucial for ensuring successful NN training, as it directly affects the model’s ability to converge efficiently and avoid training issues.

Beyond basic gradient descent, several variants have been developed to improve the efficiency and quality of NN training. These include momentum-based gradient descent, Nesterov Accelerated Gradient (NAG), and Adam (Adaptive Moment Estimation) \cite{Gower, adamopt}, among others. Gradient descent with momentum adds a ``memory'' term that accelerates movement in consistently downhill directions, while Nesterov's method applies an additional correction anticipating the future direction. Adam combines momentum accumulation and adapts the learning rate for each parameter, leading to faster and more stable convergence in many practical applications.

The gradients of the loss function in Equation \eqref{eq-loss}, with respect to $w$ and $b$, are given by the usual calculus chain rule:
\begin{align}
    \nabla_w \mathcal{L} &= -2(y_j - \hat{y}) f' \mathbf{x}_j,  \label{eq-nabla2}\\
    \nabla_b \mathcal{L} &= -2(y_j - \hat{y}) f'.
    \label{eq-nabla}
\end{align}

The derivation of Equations \eqref{eq-nabla2} and \eqref{eq-nabla}, along with examples of different loss functions, is detailed in Appendix A. In each case, the term $f'$ is determined by the chosen activation function $f$. Some derivative values for different activation functions are presented in Table 1 in the appendix. In our example, we use the sigmoid function, leading to:
\begin{align*}
    w_{\text{new}} &= w_{\text{old}} - 2 \eta(y_j - \hat{y} ) \hat{y}(1 - \hat{y}) \mathbf{x}_j, \\
    b_{\text{new}} &= b_{\text{old}} - 2 \eta(y_j - \hat{y} ) \hat{y}(1 - \hat{y}).
\end{align*}

Implementing the weight and bias update in \textit{Python}:
\textit{Python}:
\noindent \begin{lstlisting}[language=Python, caption=Updating weights and bias in Python]
eta = 0.01
# Updating weights
w -=  2*eta*( y_j -  y)*y*(1-y)*x
b -=  2*eta*( y_j -  y)*y*(1-y)
print(w,b)
# Script output
[ 1.0002,  0.000138, 0.50005,-0.29], [0.500039]
\end{lstlisting} 

After completing the learning process for the perceptron using the first dataset sample ($j=0$), the procedure is repeated for each remaining sample. Thus, each row $j$ in the dataset contributes to refining the weights. Summarizing the steps for all $M$ samples into a single script:

\noindent \begin{lstlisting}[language=Python, caption=Iterating through all samples in the dataset]
w = [1, 0, 0.5, -0.3]
b = [0.5]
eta = 0.01
Loss = 0
for j in range(M):
    x_j = dataset.data[j,:]  # Input
    y_j = dataset.target[j]  # Target/label
    z   = np.dot(w,x_j) + b 
    y   = sigmoid_activation(z) 
    w   = w - 2*eta*(y_j-y)*y*(1-y)*x
    b   = b - 2*eta*(y_j-y)*y*(1-y)
    loss += (y_j-y)**2/M
\end{lstlisting}  

This loop must be iterated multiple times (each full pass through the dataset is called an {\it epoch}) until the loss function approaches its minimum (in this case, approximately zero), indicating that the perceptron has successfully learned the provided information. To perform multiple iterations, and therefore many epochs, we add another \textit{for} loop. The complete code is available on GitHub \cite{Github}, allowing readers to reproduce the numerical experiment at their own pace.

\begin{figure}[!ht]
    \centering
    \includegraphics[width=1\linewidth]{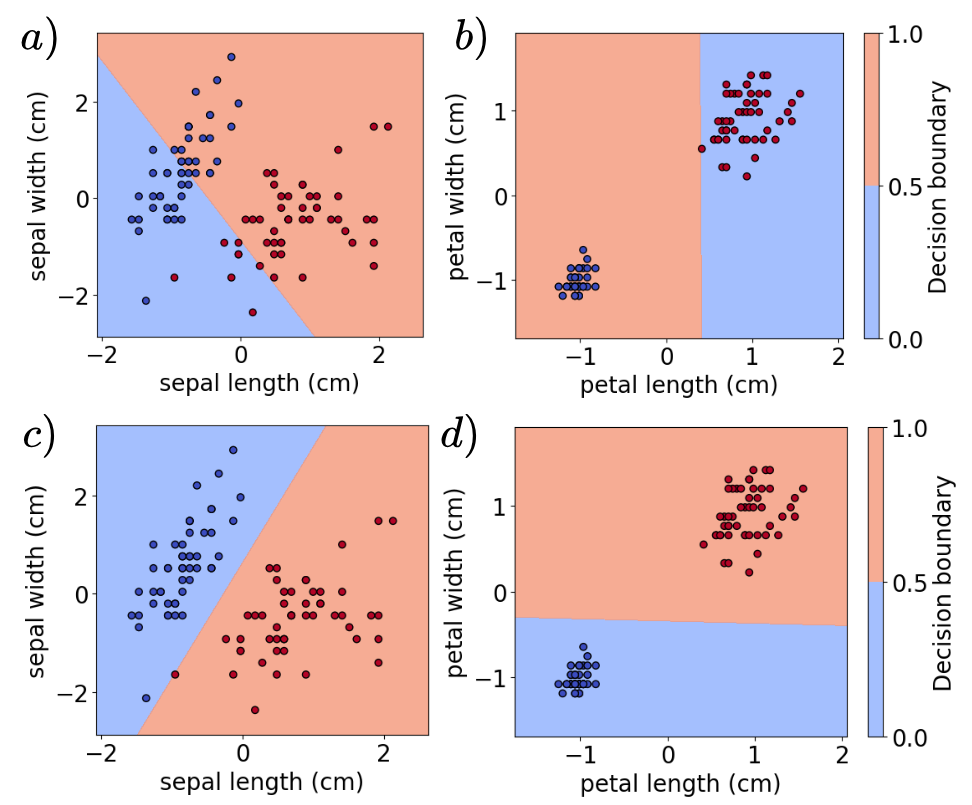}
    \caption{Evolution of the perceptron’s decision boundaries on the Iris dataset. The data points are plotted as a scatter plot projected onto two of the dataset’s four dimensions. On the left, a) and c) show the dimensions corresponding to sepal length (x-axis) and sepal width (y-axis), while on the right, b) and d) correspond to petal length (x-axis) and petal width (y-axis). In both, the learning labels are color-coded: red for class 0 and blue for class 1. The top graphs, a) and b), show the classification obtained with randomly initialized weights, whereas the bottom graphs, c) and d), show the classification after optimization. Observe that after optimization, the hyperplane (known as the \textit{decision boundary}, where the perceptron output is 0.5) separates the data clusters clearly in both projections, except for one point.}
    \label{fig:antes-depois}
\end{figure}

In Figure \ref{fig:antes-depois}, taken from the supplementary material available on GitHub, we present four two-dimensional graphs: the two at the top show the decision boundary before training (with arbitrary weight initialization), and the two at the bottom show the boundary after parameter optimization, illustrating the decision regions learned by the perceptron on the Iris dataset. In each pair of graphs, the left graph represents the decision boundary in terms of sepal length and width, while the right graph represents it in terms of petal length and width. The points correspond to dataset samples, where blue and red colors indicate the different classes the perceptron is supposed to distinguish after training. To construct the decision boundary, the NN was configured so that all activation function outputs greater than $0.5$ are classified as belonging to class (1), and conversely, outputs less than or equal to $0.5$ are classified as belonging to class (0). This setting defines a cutoff line that separates the two classes in feature space, enabling the model to distinguish between them based on computed outputs.

In graphs \ref{fig:antes-depois}.a and \ref{fig:antes-depois}.b, we observe a significant overlap of the color regions, indicating a considerable difficulty in separating the classes. This is characterized by an ill-defined decision boundary and reflects the low accuracy of the model before any learning takes place. In contrast, graphs \ref{fig:antes-depois}c and \ref{fig:antes-depois}d show a clear separation between the classes, with the colored regions now distinctly divided by a well-fitted decision boundary after 100 training epochs.

\subsubsection{Didactic Example: Linear Regression}
While the previous example addressed a classification problem, which involves categorizing data into discrete-valued classes, regression aims to predict continuous values, such as the price of a product, the weight or height of a person. In Physics problems, regression can be used to predict physical quantities such as position, velocity, or acceleration over time, serving as a fundamental tool for studying relationships between physical quantities in experiments. In this example, we will create our own dataset to illustrate the process of fitting a linear model. We assume that the dataset contains only two variables: one representing the velocity of an object under constant acceleration and another representing time instants.
\begin{align}
    \mathbf{t} &= 
    \begin{bmatrix}
        0, & 0.25, & 0.5, & 0.75, & 1
    \end{bmatrix}, \\
    \mathbf{v} &= 
    \begin{bmatrix}
       0, & 0.125, & 0.25, & 0.375, & 0.5
    \end{bmatrix}.\label{Passo0}
\end{align}

Our goal is to use the perceptron as a function that maps input data to outputs. Here, $\mathbf{t}$ represents the input data and $\mathbf{v}$ the experimental points to be fitted. Thus, we want the model to learn to predict velocity given time. With the data in hand, the next step is to generate random weights and biases, for which we use:
\begin{align}
    \mathbf{w} = [0.497], \mathbf{b} = [-0.138].\label{Passo1}
\end{align}

Next, we perform the \textit{forward} step, where input data is propagated through the model using Equation \eqref{eq3} for each value of $t$. 

Before proceeding, it is important to clarify a key concept in NNs: \textit{batching}. This technique involves dividing the dataset into small subsets called \textit{batches}. Instead of computing the loss function for each individual example, batching allows the calculation to be performed on a group of examples at once. Then, the model parameters are updated based on the average or sum of the gradients computed for all batch elements. This approach not only improves computational efficiency but also enhances generalization by reducing stochastic noise in parameter updates. Batching is widely used to accelerate model convergence and stabilize optimization, especially in large datasets. Below, we detail some of the mathematical steps in the perceptron case to help the reader understand the process in its entirety.

We begin by computing the inner product over a batch of 5 time points:
\begin{align*}
    &\mathbf{w}\mathbf{t} + b = [0.497]*[0,0.25,0.5,0.75,1] +[-0.13]\\
    &\mathbf{w}\mathbf{t} + b = [-0.138  , -0.01375,  0.1105 ,  0.23475,  0.359]. 
\end{align*}

Next, we apply the nonlinear transformation using the ReLU function (see Appendix A) to the entire batch:
\begin{align*}
    &f(\mathbf{w}\mathbf{t} + b) = f([-0.138  , -0.01375,  0.1105 ,  0.23475,  0.359]) \\
    &f(\mathbf{w}\mathbf{t} + b) = [0, 0, 0.1105 ,  0.23475,  0.359].
\end{align*}

The result $f(\mathbf{w}\mathbf{t} + b)$ represents the predicted velocity values $\hat{y}$, which we use to compute the error with the loss function in Equation \eqref{eq-loss}:
\begin{align*}
    \mathcal{L} &= \sum_{\text{batch}} (\mathbf{v} - \hat{y})^2,\\
    \mathcal{L} &= 0 + 0.0156 + 0.0194 + 0.0196 + 0.0198,\\
    \mathcal{L} &= 0.0746.
\end{align*}

To update the weights and bias, we use Equations \eqref{eq-nabla2} and \eqref{eq-nabla}. The derivative of the ReLU function follows the rule: $f'(x) = 1$ if $x > 0$, and $f'(x) = 0$ if $x \leq 0$.
\begin{align*}
    f'(\mathbf{w}\mathbf{t} + b) &= f'([0, 0, 0.1105 ,  0.23475,  0.359]) \\
    f'(\mathbf{w}\mathbf{t} + b) &= [0, 0, 1, 1, 1].
\end{align*}

Assuming that $\eta = 0.05$, we obtain:
\begin{align}
      2\eta(\mathbf{v} - \hat{y})f'_j &= [0, 0, 0.013,0.014025, 0.0141] \\
    2\eta(\mathbf{v}- \hat{y})f'_{j}\mathbf{t} &= [0, 0, 0.0069, 0.0105, 0.0141]\label{Passo3}.
\end{align}

Taking the batch average, we obtain:
\begin{equation*}
     \eta \sum_{\text{batch}} \nabla_w \mathcal{L} = 0.0063.
\end{equation*}

Using steps \eqref{Passo0}, \eqref{Passo1}, and \eqref{Passo3} in Equation \eqref{eq-nabla}, we can compute the updated weights and biases:
\begin{align*}
    w_{\text{new}}  &= w_{\text{old}} - 0.0063, \\
    w_{\text{new}}  &= 0.4906, \\
    b_{\text{new}}  &= b_{\text{old}} - 0.008, \\
    b_{\text{new}}  &= -0.1464.
\end{align*}

\begin{figure} [h]
    \centering
    \includegraphics[width=1\linewidth]{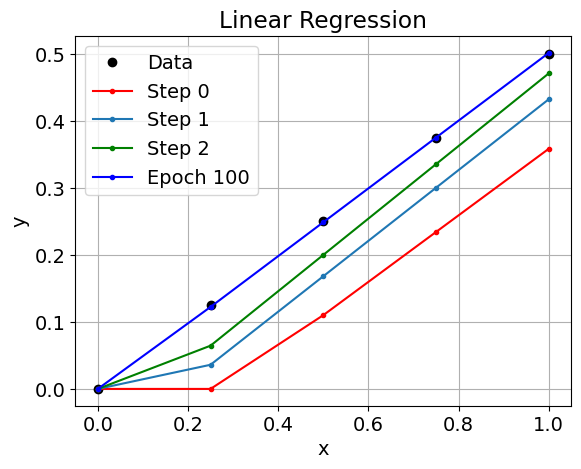}
    \caption{Linear regression fitting behavior using a perceptron over different training epochs. The black dots represent training data points, while the red, green, and blue lines correspond to the model’s predictions at epochs 0, 1, 2, and 100, respectively.}
    \label{fig:regressao}
\end{figure}

Finally we have completed the first epoch and expect to obtain a lower loss function value than before. We can verify this by repeating the previous process once more with the new weights.
\begin{align*}
    \hat{y}&=f(\mathbf{w}\mathbf{t} + b) = [0, 0.036, 0.178, 0.32, 0.46],\\
    \mathcal{L} &=\sum_{\text{batch}} (\mathbf{v} - \hat{y})^2 = 0.0643.
\end{align*} 
Comparing the new loss function value with the previous one, we observe a reduction of $0.0103$, indicating that the perceptron is learning from the data. This process can be repeated multiple times to progressively reduce the loss function. In our supplementary material on GitHub, we continue this example, running the algorithm multiple times, for a total of 100 epochs. The results of this test are shown in Figure \ref{fig:regressao}, where we achieve an error value of $1.059 \times 10^{-7}$.

In Figure \ref{fig:regressao}, we observe the convergence of the linear regression model over training epochs, with the model line progressively fitting the observed data. This behavior demonstrates the perceptron’s ability to learn and adjust to a simple dataset through successive parameter updates.

\subsection{Deep Neural Networks}
\label{sec:RNProfunda}
An artificial NN consists of a set of interconnected artificial neurons, whose disposition defines the model’s structure. The way these neurons are connected is known as the network’s architecture. There are several ways to organize these connections, and therefore many possible architectures. A common configuration involves grouping neurons into functional units known as layers. Each layer consists of neurons that operate independently but process information before transmitting their outputs to the next layer.

A well-known architecture in ML is the \textit{Multi-Layer Perceptron} (MLP), a type of dense \textit{feed-forward} NN. \textit{Feed-forward} networks are characterized by the absence of recurrent connections, meaning that outputs from one layer are used as inputs exclusively for the next layer. This unidirectional information flow goes from the input layer to the output layer. Specifically, the neurons in layer $n$ provide inputs to the neurons in the subsequent layer, $n+1$, without feedback loops or cycles in the process. This architecture is widely used due to its straightforwardness and effectiveness in classification and regression tasks, where a direct mapping between inputs and outputs is desired.

\begin{figure}[!ht]
    \centering
    \includegraphics[width=1.0\linewidth]{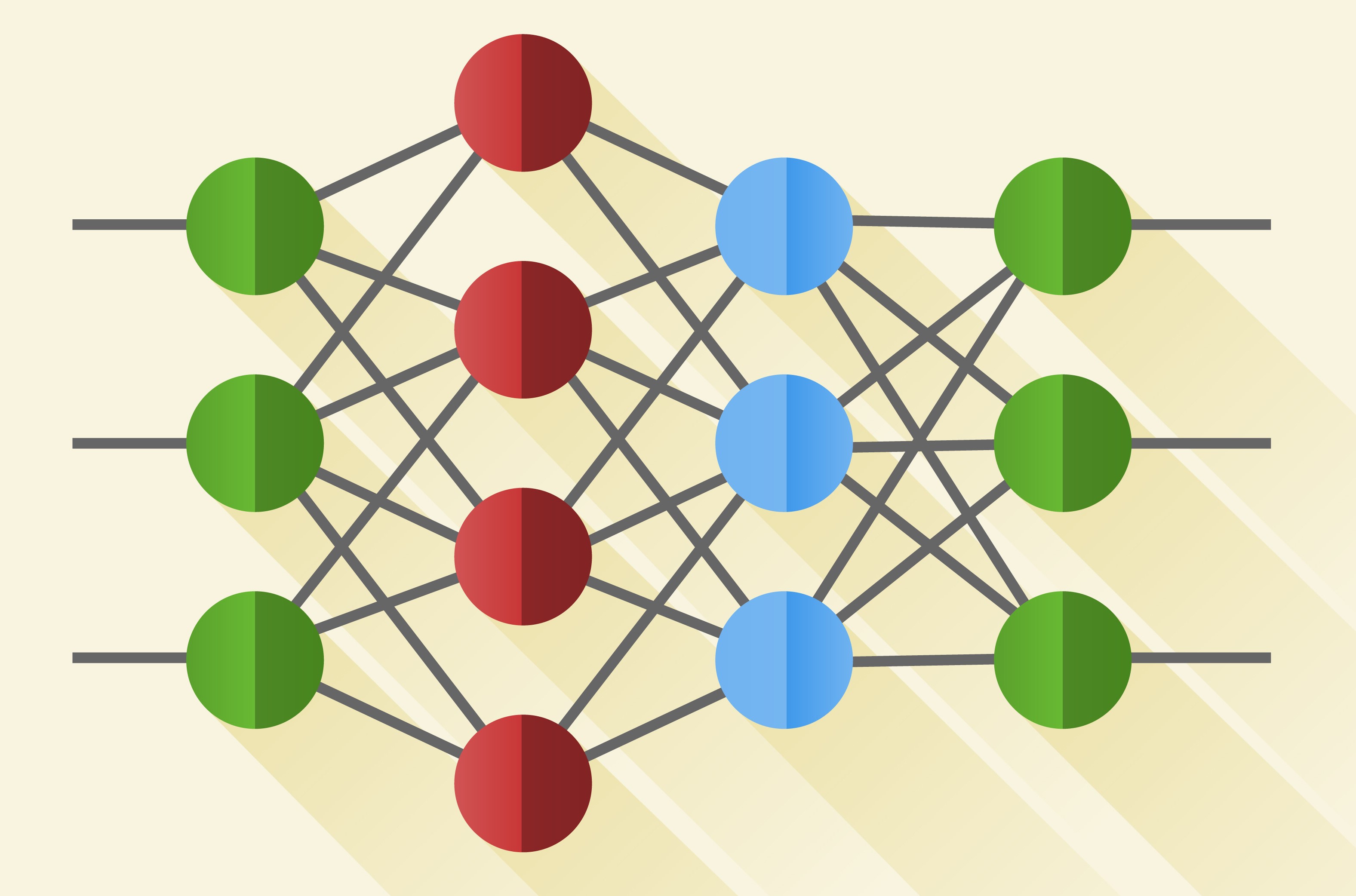}
    \caption{Pictorial representation of an NN using graphs. The green circles on the left represent the inputs, and the green circles on the right represent the outputs. The middle layers, composed of red and blue neurons, are the hidden layers. Source: Medium \cite{MediumSafrin}.}
    \label{fig:NeuralNetwork}
\end{figure}

The graphical representation of the MLP architecture is illustrated in Figure \ref{fig:NeuralNetwork}, read from left to right: the inputs are represented by the lines originating from the left edge of the image, connected to the green nodes, denoted as $x_{ji}$ in the notation of the previous chapter. The inputs are connected to the first layer (in red), where each node is a perceptron whose output serves as input for the second layer of neurons (in blue), which in turn connects to the final layer of neurons on the right (also in green). These final neurons correspond to the network’s output. Each node may have its own activation function, and the neuron-to-neuron connections can be arbitrarily determined, provided they always connect only to adjacent layers. The intermediate layers between the input and output layers (in this example: red and blue) are called \textit{hidden layers}.

For simplicity, we will use the term "neural networks" to specifically refer to artificial feed-forward NNs. However, it is important for the reader to understand that the field of NNs is quite vast and includes a wide variety of architectures, each designed with specific features to address different types of problems. Some architectures worth mentioning are: Recurrent neural networks (RNNs), which stand out as particularly effective for handling sequential data, such as time series \cite{NathanKutz2017,Rumelhart,LeCun} and natural language processing, often employing LSTM (\textit{Long Short-Term Memory}) networks \cite{Hochreiter}; Convolutional neural networks (CNNs), which we will explore in greater depth in later sections, are widely used in computer vision tasks, such as image recognition and object detection, due to their ability to extract hierarchical features from spatial data \cite{Marquardt2021};  Transformers, which have revolutionized natural language processing and have also been successfully applied in other domains, are notable for their ability to capture long-range dependencies in sequential data without requiring recurrent processing \cite{transforme}; Other architectures, including autoencoders, deep belief networks, generative adversarial networks (GANs), and Boltzmann machines combined with Bayesian deep learning methods, have significantly contributed to the advancement of the field \cite{Michelucci,Alom}. A pictorial representation of the many possible architectures can be found in \cite[257]{NathanKutz2017}. We recommend the work of Herberg et al. \cite{Herberg} for the reader who is interested in exploring these architectures in detail.

Recently, John J. Hopfield and Geoffrey Hinton were awarded the 2024 Nobel Prize in Physics for their foundational contributions to machine learning using artificial NNs. Hopfield developed networks capable of storing and reconstructing data patterns using physics-based principles, known as Hopfield Networks \cite{hopfield}, while Hinton developed the Boltzmann machine, which recognizes patterns in data using Statistical Physics tools. These innovations laid the groundwork for major advancements in artificial intelligence and deep learning.

The choice of NN architecture should be carefully considered and is closely tied to the specific problem to be solved, the characteristics of the data involved, and the model’s performance and efficiency requirements. In general, the objective of an NN is to approximate an ideal function ${g: U \rightarrow V}$, which is not necessarily known, where $U,V$ are vector spaces. Thus, given $\mathbf{x} \in U$ as an element of the dataset, and $\boldsymbol{W},\boldsymbol{b} \in W$ as parameter vectors, we aim to find a NN function ${\textit{NN}: U \times W \rightarrow V}$ that satisfies 
\begin{equation}
     \textit{NN}(\mathbf{x}; \boldsymbol{W},\boldsymbol{b}) \approx g(\mathbf{x}).
\end{equation}

\noindent In a regression problem we may have an input $\mathbf{x}$ associated with a value $g(\mathbf{x})$, where the NN should output the function $g(\mathbf{x})$ value, through optimizing the values of the parameters $\boldsymbol{W}, \boldsymbol{b}$, resulting in the best approximation of this function, $\textit{NN}(\mathbf{x}; \boldsymbol{W},\boldsymbol{b})$.

One can mathematically describe the \textit{feed-forward} architecture, considering fully connected neurons between layers, as pictorially represented in Figure \ref{fig:NeuralNetwork}. The network transforms the input \(\mathbf{x}\) into an output \( \textit{NN}(\mathbf{x}; \boldsymbol{W},\boldsymbol{b})\) using the following structure (considering the first layer as red in the figure and the input as the green nodes).

\begin{align*}
    \textbf{ Input }&= \mathbf{x} \\
    \textbf{1st Layer }&= \Big(f^{[1]}_1 ( \sum_{i=1}^{N_0}{w^{[1]}_{i1} x_i} + b^{[1]}_1), \ldots, \\  &\qquad \ldots, f_{N_1}^{[1]}(\sum_{i=1}^{N_0} w^{[1]}_{iN_1} x_i + b^{[1]}_{N_1}) \Big)\\
    \textbf{ 2nd Layer }&= \Big( f^{[2]}_1 ( \sum_{i'=1}^{N_1} w^{[2]}_{i'1} f^{[1]}_{i'} (\ldots) + b^{[2]}_1   ), \ldots \\
    &\qquad \ldots,   f^{[2]}_{N_2} ( \sum_{i'=1}^{N_1} w^{[2]}_{i'N_2} f^{[1]}_{i'} (\ldots) + b^{[2]}_{N_2}   ) \Big)\\ 
     &~\vdots  \\
    \textbf{ Output }&= \Big( f^{[n]}_{1} (\sum_{i'=1}^{N_{(n-1)}} w^{[n]}_{i'1} f^{[n-1]}_{i'} (\cdots) + b^{[n]}_{1}   ), \ldots, \\
    & \qquad \ldots, f^{[n]}_{N_n} (\sum_{i'=1}^{N_{(n-1)}} w^{[n]}_{i'{N_n}} f^{[n-1]}_{i'} (\cdots) + b^{[n]}_{N_n} ) \Big) \\
    &= \textit{NN}(\mathbf{x}; \boldsymbol{\theta})  
\end{align*}
where $f^{[n]}_{k_n}$ and \(b^{[n]}_{k_n}\) are the activation function and bias vector of the $n$-th layer associated with the $k$-th neuron of that layer, respectively, and \(w^{[n]}_{lk_{n}}\) is the weight that connects the $l$-th neuron of layer $n-1$ to the $k$-th neuron of layer $n$.

The reader might be asking herself if one can be sure that this long chain of iterative composition of weighted sums and activations actually approximates the desired function $g(\mathbf{x})$. The answer to this question is affirmative, given the following theorem \cite[p.36, p.3]{dawid_modern_2022, KAN2024}.

\begin{theorem}[Universal Approximation (Kolmogorov-Arnold)]
    Let $g: \mathbb{R}^N \rightarrow \mathbb{R}$ be a continuous function, and let $\varphi_q,\phi_{q,p}:\mathbb{R} \rightarrow\mathbb{R}$ be nonlinear functions, with $q\in \{0,\ldots,2N\}$ and $p \in \{0,\ldots,N\}$. Then $g$ can be approximated by 
    \begin{equation}
        f(\mathbf{x}) = \sum_{q=0}^{2N} \varphi_q\Big(\sum_{p=0}^N \phi_{q,p}(x_p)\Big) = g(\mathbf{x})
    \end{equation}
    where $\mathbf{x} \in \mathbb{R}^N$.
\end{theorem}

\noindent This means that any continuous function can be represented with a polynomial number $\mathcal{O}(N^2)$ of single-variable nonlinear functions. The reader may have noticed that the function in the theorem is similar to a two-layer NN with $b^{[n]}=0, \forall n$. In fact, with just two layers, any continuous function can be approximated, provided there are enough neurons. Pragmatically, in the practice of ML, it is preferable to use a larger number of smaller layers to improve computational efficiency \cite[p.5]{Marquardt2021}.

Considering the previous theorem, one might argue that ML and NNs are \textit{merely} methods for curve fitting and data approximation. This statement is not incorrect, but it is an oversimplification. Consider the following analogy: ``Quantum many-body physics is just the Schrödinger equation in high-dimensional spaces''. While technically true, we know that in practice, there are many emerging phenomena in this field that require new techniques absent in few-body quantum mechanics, making the statement overly reductive. The same can be said for NNs \cite[p.3]{Marquardt2021}.

For practical applications, constructing deep NNs relying only on libraries like \textit{NumPy}, as we did for the perceptron in Section \ref{sec:perceptron}, is a overwhelming task. Due to the complexity and size of NNs, it is preferable to use ready-made Python libraries such as \textit{PyTorch} \cite{pytoch}, \textit{TensorFlow} \cite{tensorflow}, and \textit{Jax} \cite{jax}, to simplify usage, as many typical NN training operations (which we discussed earlier) come as built-in functions in such libraries. Moreover, a fundamental feature of these libraries is automatic differentiation, which, due to its embedded structure, is essential to improve NN training. This technique efficiently computes gradients precisely and automatically, eliminating the need for manual differentiation or numerical approximations such as finite differences \cite{autodiff}. Automatic differentiation automates these calculations using the standard chain rule for derivatives, facilitating the application of algorithms such as gradient descent. This accelerates development, ensures greater accuracy, and preserves differentiable structures, making these tools widely adopted in ML.

The reader can find additional material for this article on our GitHub repository at \cite{Github}, where we provide \textit{Jupyter notebooks} with step-by-step explanations on how to create an NN using the \textit{PyTorch} library. To introduce the library, we have prepared a specific \textit{Jupyter notebook} applied to the same flower classification problem (\textit{Iris dataset}) solved earlier.
\begin{figure}[!ht]
    \centering
    \includegraphics[width=1\linewidth]{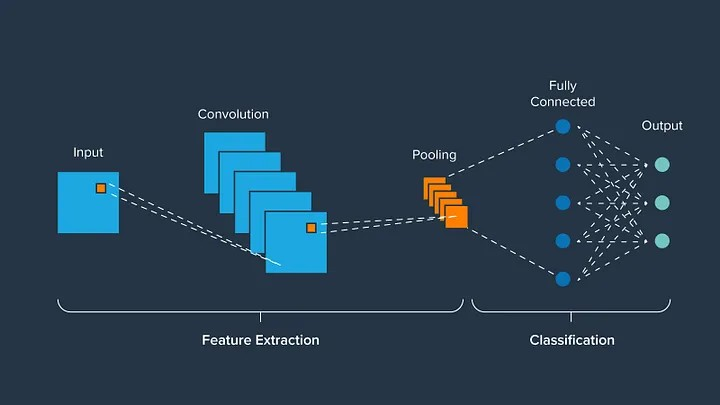}
    \caption{Pictorial representation of a Convolutional Neural Network (CNN). Given an image (a matrix), submatrices (in yellow) of this data serve as input to the filter layer. Each filter performs a convolution over these submatrices, with each kernel specialized in extracting a specific feature from the image. The result of this layer is processed by a pooling layer, which reduces the dimensionality of the resulting matrices. Finally, the reduced matrices are passed on to a fully connected layer that learns patterns from these lower-dimensional representations. Source: Medium Khang \cite{MediumKhangPham}.}
    \label{fig:cnn}
\end{figure}

\textbf{Convolutional Neural Networks (CNNs)}  
Convolutional Neural Networks (CNNs) differ from fully connected networks by two fundamental concepts: the locality of connections and parameter sharing. Locality refers to the fact that a neuron, or a set of neurons, is connected only to a subgroup of neurons in the previous layer, rather than being connected to all the neurons in that layer, as in fully connected networks. This feature allows CNNs to capture local patterns, such as edges and textures in images, while maintaining a more compact representation of the data.

Parameter sharing, on the other hand, means that all convolutional neurons within the same layer, called {\it filters}, use the same set of parameters, including weights and biases. These neurons are distinguished only by the input subspace to which they are connected, enabling efficient detection of similar features in different parts of the input, such as an image. Due to this property, CNNs are highly efficient for tasks where the relative position of a feature is more important than its exact location, such as in pattern recognition and computer vision \cite{bojarski2017,NIPS2012_c399862d}.

Moreover, it is common in convolutional architectures to apply multiple filters in parallel over the same input values. Each filter has its own set of parameters, which are shared among the neurons it consists of. This approach allows the network to extract different types of features, such as horizontal edges, vertical edges, or complex textures, simultaneously.

When one uses the term `convolution' in Physics, one is generally expressing the idea that if a particle at $x$ is functionally represented by $f(x)$, and the way it interacts with its environment (point by point) is governed by a function (or a field) $G(x)$, then the interaction of the surroundings (say, at neighboring points $x'$) with the particle can be represented by the convolution $A(x) = \int G(x-x')f(x')dx'$ \cite[p.13]{Marquardt2021}. In general, the function $G(x-x')$, called the \textit{kernel}, preserves a number of properties that are known to be important in physics, such as causality. The same idea is applied in CNNs: The kernel of the convolution, the \textit{filter}, provides information about how a pixel in an image is related to its neighbors. The term `filter' was inherited from the computer vision research field \cite{Lindeberg1998}.

\begin{figure} [ht]
    \centering
    \includegraphics[scale=0.25]{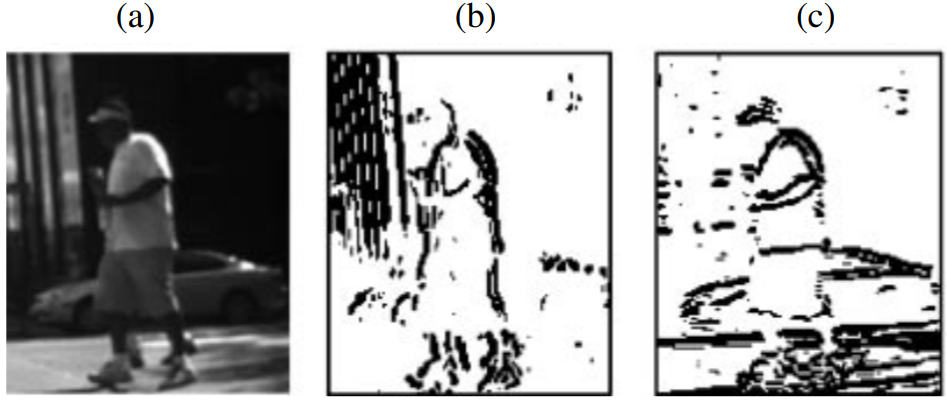}
    \caption{Examples of convolution uses for edge detection. (a) Original image, (b) under the filter $G_x$, (c) under the filter $G_y$. Filters are explicitly expressed in Equation (\ref{eq:ExempKernel}). Source \cite[p.4]{EdgeDetec2008}  }
    \label{fig:FiltrosCNN}
\end{figure}

See in Equation (\ref{eq:ExempKernel}) two examples \cite{EdgeDetec2008} of filters used to detect edges in images, in the $x$ and $y$ directions, and their effects on images in Figure \ref{fig:FiltrosCNN} (these filters are known in computer vision as {\it edge detection filters} or {\it Sobel filters} \cite{LindebergScaleSpace1993}).

\begin{equation}
    G_x = \frac{1}{3}
    \begin{bmatrix}
    -1 & 0 & 1 \\
    -1 & 0 & 1 \\
    -1 & 0 & 1 \\
    \end{bmatrix}, \qquad 
    G_y = \frac{1}{3}
    \begin{bmatrix}
    -1 & -1 & -1 \\
    0 & 0 & 0 \\
    1 & 1 & 1 \\
    \end{bmatrix}
    \label{eq:ExempKernel}
\end{equation}

In the case of CNN (as opposed to deterministic computer vision), filters are not predefined, but are instead initialized randomly. Instead, during the loss function optimization process, the filter parameters are updated and converge to effective filters. That is, although there are many deterministic filters available in computer vision, in general, the network should be capable of generating its own filters that minimize the cost function, which can often be non-trivial.

It is common practice in the literature for convolutional layers to be followed by pooling layers and subsequently dense layers. This mixed architecture is represented in Figure \ref{fig:cnn}. The pooling layers play a role in reducing the spatial dimensions of the input while preserving the most important features of the image or input data. This dimensional reduction makes the neural network more computationally efficient and less sensitive to small variations in the position of the detected features. The dense layers at the end serve the purpose of deducing from the analyzed images the desired effect, whether it be classification, noise removal, etc. \cite[247]{NathanKutz2017}. There is a biological inspiration for this choice, but it is beyond the scope of this work and can be found in \cite{FukushimaCNN1980}.

Pooling is an operation that considers a set of elements, such as a matrix of size $K \times K$, and applies a reduction rule that generates a smaller matrix, for example, $\frac{K}{2} \times \frac{K}{2}$ \cite[p.247]{NathanKutz2017}. There are several ways to perform this reduction, each with its own characteristics and advantages. Let us consider two examples:

1) Given a block of pixels from a larger image, $B_{4 \times 4}$, i.e., a submatrix of the input data matrix, then

\begin{align*}
    &\textit{pooling} (B_{4\cross 4}) = M_{2 \cross 2}, \\    
    &\text{where } M_{ij} = \frac{1}{4} \sum_{k=i,l=j}^{i+2,j+2} B_{kl}.
\end{align*}

\noindent i.e., each element $M_{ij}$ is the average of the elements of a $2 \times 2$ submatrix of $B_{4 \times 4}$.

2) From the same block of pixels $B_{4 \times 4}$, taking the largest of the elements of a $2x2$ submatrix $M_{ij} = \max_{k=i,l=j}^{i+2,j+2}(B_{kl}) $, provides a different kind of pooling. Note that after such an operation, the size of the processed input data is now compressed into a smaller matrix. Reducing the dimensionality of the data decreases the network's processing time and thus speeds up the learning process as well. However, loss of information is always present in these processes, so a balance between the two operations must be found according to contextual use.

\subsection{Training in Deep Networks}

When dealing with deep networks, we will see that the training algorithm, in its essence, preserves the same rules as those used to arrive at equations (\ref{eq:SGD_w_perceptron}) and (\ref{eq:SGD_b_perceptron}), but some care must be taken. Note that now the weights and biases require an additional index compared to the perceptron case, which indicates the layer from which the weight originates. We want to optimize the weights of any arbitrary layer, but we only know the result of the cost function, which is calculated by the network's output, and this, in turn, is a composition of all the internal operations in the network, in a well-defined order. Fortunately, there is a well-known tool for determining the derivative of composite functions, which is the chain rule. Recall that the output of the $l'$-th neuron $y^{[n]}_{l'}$ in the $n$-layer network has a derivative with respect to a weight on the $k$-layer $w^{[k]}_{il}$, $\frac{d}{dw^{[k]}_{il}} y^{[n]}_{l'} $. Writing the result of the weighted sum $z^{[n]}_{l'}= \sum_{i'=1}^{N_{(n-1)}} w^{[n]}_{i'l'}  y^{[n-1]}_{i'}+ b^{[n]}_{l'} $, i.e., calculated in the $l'$-th neuron of the $n$-th layer, we have, by the chain rule,

\begin{align*}
    \frac{d}{dw^{[k]}_{il}} y^{[n]}_{l'} &= \frac{d}{dw^{[k]}_{il}} \Big( f^{[n]}_{l'} (z^{[n]}_{l'})\Big)   \\
    &=\frac{d}{dz^{[n]}_{l'}} f^{[n]}_{l'} (z^{[n]}_{l'})   \cdot \sum_{i'=1}^{N_{(n-1)}}  w^{[n]}_{i'l'} \frac{d}{dw^{[k]}_{il}} \Big( y^{[n-1]}_{i'} \Big) 
\end{align*}

\noindent Therefore, the derivative $\frac{d}{dw^{[k]}_{il}} y^{[n]}_{l'} $ depends on the derivative of $\frac{d}{dw^{[k]}_{il}} y^{[n-1]}_{i'} $, which shows that we have a recursive structure in this calculation. Moreover, since the indices $i'$ repeat in the sum $$\frac{d}{dw^{[k]}_{il}}z^{[n-1]}_{l'} = \sum_{i'=1}^{N_{(n-1)}}  w^{[n]}_{i'l'} \frac{d}{dz^{[n-1]}_{i'}} f^{[n-1]}_{i'} (z^{[n-1]}_{i'}) \frac{d}{dw^{[k]}_{il}} \Big( z^{[n-1]}_{i'}\Big),$$ we can define a matrix

\begin{equation}
    M^{[n,n-1]}_{i',l'} = w^{[n]}_{i'l'} \frac{d}{dz^{[n-1]}_{i'}} f^{[n-1]}_{i'} (z^{[n-1]}_{i'}) 
\end{equation}

\noindent Thus, we can write, recursively,

\begin{equation}
\frac{d}{dw^{[k]}_{il}}z^{[n-1]}_{l'} =  M^{[n,n-1]}M^{[n-1,n-2]}\ldots M^{[k+1,k]}\frac{d}{dw^{[k]}_{il}}z^{[k]}_{l}   ,
\label{eq:matrizcompositionbackprop}
\end{equation}

\noindent Note that this product is a standard matrix multiplication, so the training of the neural network is performed by a linear algorithm (given that derivative values are known), even though the network itself is non-linear. The rightmost term in the product is given by

\begin{align*}
    \frac{d}{dw^{[k]}_{il}}z^{[k]}_{l}   &= \sum_{i'=1}^{N_{k}} \frac{d}{dw^{[k]}_{il}} \Big( w^{[k]}_{i'l'}  y^{[k-1]}_{i'}\Big) \\
    &= \sum_{i'=1}^{N_{k}} \delta_{i'i}\delta_{l'l} y^{[k-1]}_{i'} \\
    &= y^{[k-1]}_{i}.
\end{align*}

\noindent Similarly

\begin{align*}
        \frac{d}{db^{[k]}_l}z^{[k]}_{l}   &= 1.\\
\end{align*}

\noindent This process is called \textit{backpropagation}. The fact that backpropagation can be represented by matrix multiplications, which can simply be stored in memory when we test a value of the network (during the `\textit{forward pass}'), and that we can then reuse these stored values for many parameters $w^{[k]}_{il}$ in parallel, is what makes the task of training networks computationally feasible in a timely manner. If \textit{backpropagation} were not efficient, none of the neural network applications today would be possible. Next, we will see a step-by-step explanation of the \textit{backpropagation} algorithm inspired by \cite{Marquardt2021}, in the same way we did for updating the weights of the perceptron in Section \ref{sec:perceptron}.

\begin{enumerate}
    \item Start by calculating the deviation of the \textit{loss} function from the true value with respect to an output $l$ and its derivative but applying the chain rule only once.
    \begin{equation}
        \Delta_{l'} = (y^{[n]}_{l'} - \hat{y}_{l'})\frac{d}{dz_{l'}^{[n]}}f_{l'}(z^{[n]}_{l'}) 
    \end{equation}
    \item Now, since the derivatives $\frac{d}{dz_{l'}^{[k]}}f_{l'}(z^{[k]}_{l'})$ were calculated during the \textit{feedforward} step where we computed $f_{l'}(z^{[k]}_{l'})$, retrieve them from memory and update delta with the matrix composition $\hat{M}$ from the equation. (\ref{eq:matrizcompositionbackprop})
    \begin{equation}
        \Delta_{l'} \leftarrow \Delta_{l'}\hat{M}\frac{d}{dw^{[k]}_{il}}z^{[k]}_{l} 
    \end{equation}
    \item Repeat steps 1 and 2 for all $l'$. Now take the average of the resulting deltas $\Delta$, and update the weight $w_{il}^{[k]}$ as we did in the equation.(\ref{eq:SGD_w_perceptron}) e (\ref{eq:SGD_b_perceptron})
    \begin{align}
        &w_{il,\text{new}}^{[k]} \leftarrow w_{il,\text{old}}^{[k]} - \eta\Delta \\
        &b_{l,\text{new}}^{[k]} \leftarrow b_{l,\text{old}}^{[k]}- \eta\Delta 
    \end{align}
\end{enumerate}

The \textit{backpropagation} algorithm has limitations that can compromise its performance and applicability in deep NNs. One of the main limitations is the problem of exploding and vanishing gradients, which occurs when the gradients calculated by the algorithm are excessively large or small \cite[p.142]{NathanKutz2017}. This phenomenon can result in the cumulative amplification or attenuation of gradients across layers, making it difficult for the network's parameters to converge properly, as the step size of the optimization depends on the magnitude of the gradient, which, when small, leads to slow convergence. This requires many training cycles, increasing the time and computational resources needed. Another significant limitation is that optimization algorithms using \textit{backpropagation} are strongly influenced by the initialization of parameters. This factor can result in unsatisfactory performance or the inability to find a local optimum.

Additionally, the \textit{backpropagation} algorithm tends to be more effective when used with large datasets, as it allows the NN to better capture the complexities and variations of patterns present in the problem. However, the requirement for large volumes of data can become an obstacle in domains where data collection is limited or involves ethical and privacy concerns, such as in medical or security applications. In these situations, the scarcity of data can compromise the network's ability to learn robust representations \cite{GradAmplif2020}. Furthermore, in the absence of appropriate regularization techniques, such as \textit{dropout} or normalization, NNs can easily suffer from overfitting \cite{Bashir}. This phenomenon occurs when the model excessively fits the training data, capturing both relevant patterns and dataset-specific noise, resulting in unsatisfactory performance on new, unseen data due to its limited generalization capability.

\section{Application of neural networks in physics}
\label{sec:Capitulo3}

In this chapter, we present four distinct approaches that utilize neural networks applied to the study of the simple pendulum, a classical physical system. Each approach explores different aspects of the problem, employing machine learning techniques to model, predict, and interpret the system's behavior.  

In the first approach (Section \ref{sec:EX:Pendulo}), we use supervised learning to estimate a physical parameter of the system, specifically the acceleration due to gravity \( g \), based on simulated data. Next, in Section \ref{sec:EX-EDO}, we apply deep NNs to solve the ordinary differential equation that describes the pendulum's motion, exploring the capability of these networks to approximate complex solutions of differential equations.  In the third approach (Section \ref{sec:EX:AutoencoderPendulo}), we employ autoencoders to discover the system’s latent space, aiming to reduce the dimensionality of the data and find a compact and efficient representation of the pendulum’s state. Finally, in Section \ref{sec:EX:SINDY}, we use the SINDy (Sparse Identification of Nonlinear Dynamics) method, as developed by \cite{SINDyAutoencoder_steven2019}, to identify the underlying differential equation governing the system, using images as input.  

Each of these approaches illustrates the potential of neural networks in different contexts of physical modeling, highlighting the versatility of these tools for efficiently solving complex problems.

\subsection{Example 1: Pendulum}
\label{sec:EX:Pendulo}

We begin with a fundamental example: learning parameters applied to the simple pendulum case. We chose this problem for two main reasons: The first is that the pendulum model is widely known among undergraduate students, making it an accessible starting point for introducing ML concepts in the context of physical systems. The second reason is that students are likely already familiar with parameter regression problems, which are common in educational laboratory experiments, such as estimating physical constants.  

In this example, presented in Jupyter Notebook 03, we aim to demonstrate an alternative approach to solving a well-known problem by using NNs to estimate system parameters. Although this is a \textit{toy problem}—a simplified problem designed for illustration—it helps establish the conceptual and methodological foundation, paving the way for more innovative and advanced applications that will be discussed in the following sections.  

The differential equation of the pendulum is given by:
\begin{equation}
    \frac{d^2\theta}{dt^2} + \frac{g}{\ell}\sin{\theta} = 0.
    \label{Eq:PenduloRaw}
\end{equation}

\noindent In the small-angle approximation, we know that ${\sin{\theta} \approx \theta}$, and thus the equation simplifies to
\begin{equation}
    \frac{d^2\theta}{dt^2} + \frac{g}{\ell}\theta = 0 ,
    \label{Eq:PenduloSmallAngle}
\end{equation}

\noindent with a known solution,
\begin{equation}
    \theta(t) = \theta_0 \cos{\Big(\sqrt{\frac{g}{\ell}}t}\Big).
    \label{pinn-pendulo-solution}
\end{equation}

In the context of teaching laboratories, the student likely encountered the expression for the period $ T = \frac{2\pi}{\omega} = 2\pi\sqrt{\left(\frac{\ell}{g}\right)} $, isolated the term $ g $, experimentally measured $ T $ and $ \ell $, and determined the value of $ g $.  
In this example, suppose that the experimental data given by $ data_i = \theta_i$ were obtained, for instance, through filming the pendulum at time instant $ i $, or by capturing its projection on a linearly moving tape with $ t $, like a seismograph \footnote{This works approximately if $ \theta $ is small, since the variation of $ {y(t) = -\ell(1 - \cos{\theta}) \approx \frac{\ell}{2}\theta(t)^2 } $ is of second order}. Additionally, assume that we measured $ \ell $, i.e., we also know $ \theta_0 = \theta(0) $.  

In the developed notebook, we implemented a NN that takes as input a partitioned time vector over a defined interval, represented as $ {\mathbf{t} = (t_0, t_1, \ldots,t_i,\ldots, t_M)} $. The output of the network is the predicted value for the acceleration due to gravity, denoted by $ {\textit{NN}(\mathbf{t}; \mathbf{w},\mathbf{b}) = g'} $.  

Once the NN is declared, we need to construct a cost function. This can be done by expressing the resulting function by the composition of the cosine over $ g' $, giving the approximated angular displacement as $ {\theta'(t) = \theta_0\cos{\sqrt{\left(\frac{g'}{\ell}\right)}t}} $. The cost function is then built to compare the experimental values of angular displacement, $ \theta $, with the values of $ g' $ predicted by the NN. One can then compare the error between the observed data $ \theta(t) $ and the values modeled by the equation of motion $ \theta'(t) $, with the function

\begin{equation}
    \mathcal{L} = \frac{1}{M}\sum_{i=0}^{M}(data_i - \theta'(t_i))^2.
     \label{lossdata}
\end{equation}

When $\mathcal{L} \approx 0$, we know that $g \approx g'$, and thus the NN has well approximated the desired constant, and the curve generated by $\theta'(t)$ is close to the original one. The purpose of this example is to show that a NN can be used for regression and recovering curve parameters, i.e., this approach is generalizable to more complex examples, where one has explicit knowledge of the analytical expression that generates the curve but has many data points.

\begin{figure}[!ht]
    \centering
    \includegraphics[width=1\linewidth]{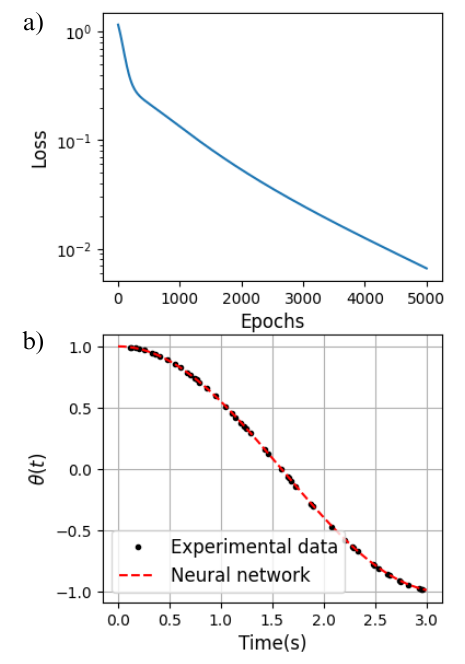}
    \caption{Result of Example 1. a) Evolution of the cost function (in logarithmic scale) over the training epochs. b) Solution for the simple pendulum, where the black dots represent the data used in the training, while the red line corresponds to the curve fitted by the neural network. }
    \label{fig:exemplo1-resultados}
\end{figure}

Figure~\ref{fig:exemplo1-resultados}, shows two graphs obtained after training the NN with only 1 neuron. The graph in Figure~\ref{fig:exemplo1-resultados}~a) represents Equation \eqref{lossdata} for each epoch, which informs us about the progress of the NN training, as decreasing values of the cost function are an indicator of the model's convergence toward a better result. However, this is not always true, due to the phenomenon of overfitting, where the NN fits the training data so closely that it loses the ability to generalize and make accurate predictions on unseen data.

Moreover, the behavior of the cost function over as a function of epochs can provide insights into the model's performance and convergence. A smooth and consistent decline in the cost function over the epochs suggests that the model is progressing steadily and learning efficiently. However, sudden fluctuations or a stagnation in the reduction of the cost function may indicate the need for adjustments in the model's hyperparameters or even the presence of issues such as overfitting. Therefore, analyzing the graph of the cost function as a function of epochs is common practice in the learning phase improving decision making and leading to better model optimization. There are other relevant problems that arise during training, such as mismatch between the cost function \cite{Huang-mismatch}, plateaus and/or stagnant cost function \cite{Ainsworth}, gradient explosion and vanishing gradients \cite{Hanin}, as well as overfitting and underfitting of models \cite{Bashir}, but these possible problems will not be discussed in depth on this text.

In Figure \ref{fig:exemplo1-resultados} b), we have a comparison of the NN's output with the theoretical expression \eqref{pinn-pendulo-solution} for the pendulum's solution, showing good agreement between the curves. This example can be found in the \textit{Jupyter Notebook} titled "03-Example 1," where the reader will find more technical details about the implementation.

\subsection{Example 2: Solving the Differential Equation}
\label{sec:EX-EDO}

In this section, we present how to solve ordinary differential equations (ODEs) using the Physics-Informed Neural Networks method \cite{PIML-review-nature}. NNs are known for their ability to approximate complex functions, which justifies their use for approximate solutions of ODEs. This application is the core of the PINN method, which integrates specific physical knowledge into the network's learning process, as detailed in the introduction of this article. To incorporate the differential equation into the network's learning, we integrate Equation \eqref{Eq:PenduloSmallAngle} into the cost function itself during training. This new function is composed not only of the differential equation but also of the initial and boundary conditions of the problem, as expressed in the following equation:
\begin{equation}
   \mathcal{L} = \mathcal{L}_{\text{data}} +\mathcal{L}_{\text{EDO}} +\mathcal{L}_{\text{IC}} + \mathcal{L}_{\text{BC}}.
    \label{Loss-pinn}
\end{equation}

Here, $\mathcal{L}_{\text{data}}$ is used when we have experimental data representing part of the solution to the equation, similar to Equation \eqref{lossdata}; $\mathcal{L}_{\text{EDO}}$ reflects the contribution of the differential equation itself, $\mathcal{L}_{\text{IC}}$ corresponds to the contribution of the initial conditions (IC), and $\mathcal{L}_{\text{BC}}$ to the boundary conditions (BC). 

To illustrate this method, we will apply PINN to the simple pendulum equation with small angles, which results in a simple harmonic oscillator. The NN, denoted as $\textit{NN}(\mathbf{t}; \mathbf{w},\mathbf{b})$, will be trained to approximate $\theta(\mathbf{t})$. Thus, the solution to the pendulum's differential equation, which will be incorporated during training, takes the following form:
\begin{equation}
    \mathcal{L}_{\text{EDO}} = \sum_i \left( \frac{d^2}{dt^2}\textit{NN}(\mathbf{t}_i; \mathbf{w},\mathbf{b}) + \frac{g}{\ell}\textit{NN}(\mathbf{t}_i; \mathbf{w},\mathbf{b}) \right)^2
\end{equation}

\noindent Since this is the differential equation for the harmonic oscillator, its ideal minimum value should be zero.

The initial condition is given solely by $\theta_0$, and it is modeled in the cost function as:
\begin{equation}
    \mathcal{L}_{\text{IC}} = \left( \textit{NN}(t=0; \mathbf{w},\mathbf{b}) - \theta_0 \right)^2
\end{equation}

\begin{figure}[!ht]
    \centering
    \includegraphics[width=1\linewidth]{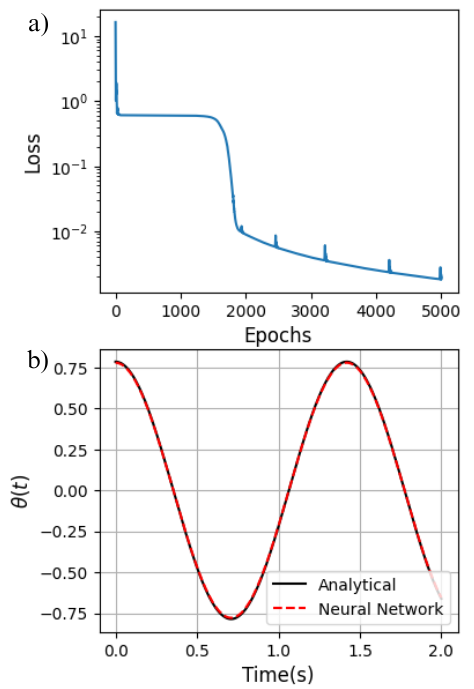}
    \caption{Result of Example 2 - solving the differential equation using neural networks with the PINN method. In a), the graph of the cost function in logarithmic scale as a function of epochs illustrates the training process, showing how the error evolves with each iteration. In b), the analytical solution of the simple pendulum is shown in black, and the result of the neural network after training is shown in red.}
    \label{fig:pinn-edo-resultados}
\end{figure}

With this approach, PINNs enable an efficient and direct integration of physical laws into the deep learning architecture, because the network now contains knowledge about the physical system, and the $\textit{NN}$ serves as an approximation of the solution to this model, considering that it satisfies the characteristic ODE by definition. Additionally, since the learning process is more structured (i.e., there is more information in the cost function), the training phases of PINNs are generally shorter and require less data. For instance, compare the graphs a) in Figures \ref{fig:exemplo1-resultados} and \ref{fig:pinn-edo-resultados}. It is notable that, despite the cost function in (\ref{Loss-pinn}) having more positive terms than the one defined in (\ref{lossdata}), the value of $\mathcal{L}\approx 10^{-2}$ is achieved in half the number of epochs compared to the case in Example~1.

In Figure \ref{fig:pinn-edo-resultados}, we have two graphs obtained after training the NN to solve the differential equation. The graph on the left (a) represents Equation \eqref{Loss-pinn} for each epoch, while in (b), we have a comparison of the NN's output with the theoretical expression \eqref{pinn-pendulo-solution} for the pendulum's solution, showing good agreement between the curves. This example can be found in the \textit{Jupyter Notebook} titled "04-Example 2," where the reader will find the technical details of the implementation.

\subsection{Example 3: Revisiting the Pendulum with Autoencoders}
\label{sec:EX:AutoencoderPendulo}

NNs can also be applied to routine tasks in computing, producing new approaches that may reveal unexpected properties of a system or offer improvements in terms of efficiency and performance. One example is the use of NNs for data compression, employing the encoder-decoder architecture, known as \textit{Autoencoder}. The \textit{Encoder} is responsible for transforming an input (such as an image) into a representation of reduced dimensionality, compressing the information into a more compact format (such as a list with fewer bits). This technique not only reduces the size of the data but can also capture the main features and abstractions of the original dataset, which can be useful for tasks such as noise removal, generation of new data, or even unsupervised learning of latent representations \cite{Lusch_KoopmanAutoencoders2018}.

Consider the following example: a curve defined by the function $f(t) = \sin{(\omega_1 t)} + \sin{(\omega_2 t)}$ has a functional expression that can become quite complex, especially when more frequencies are added to the model. However, despite the curve being composed of infinite points over time, all its information can be completely determined by knowing just two key variables: the frequencies $\omega_1$ and $\omega_2$ (here we are assuming that the phase is known and $\phi_i = 0$, for $i=1,2$). This implies that, to store the entire curve, it is sufficient to know that it is a sum of two sine functions and their respective frequencies, represented by two real numbers, illustrating how a parametric description can be conveniently compact.

Similarly, fractals exemplify another case of emergent complexity from simple rules: highly complex and visually rich figures can be generated through relatively simple iterative equations. These examples illustrate how complex information can be compactly represented by a reduced set of parameters or instructions \cite[p.324]{Gamelin2001}.

Another example of \textit{encoder} application is in the compression of black-and-white images, where pixel representation is given by a two-dimensional (2D) matrix. When an image undergoes the discrete Fourier transform, the spatial information of the pixels is converted into a frequency-domain representation. In this process, most of the significant information in the image can be captured by a relatively small set of frequency coefficients (this is known as applying a low-pass filter \cite{NathanKutz2017}). Thus, instead of storing all pixel values, it is possible to represent the entire image compactly, using only the most relevant frequencies that compose its structure.

A well-known algorithm that compresses data using the Fourier transform is the JPEG image format \cite[p.88]{NathanKutz2017}. Given compressed data, to access the original version, an algorithm that performs the inverse operation is necessary, which, taking the compressed data, recovers (even if partially) the initial image. The algorithm that performs this task is called the \textit{Decoder}, and, in the case illustrated above, it would be the inverse Fourier transform.

By applying both algorithms in a 
concatenated manner, \textit{encoder} followed by \textit{decoder}, we obtain a resulting algorithm called $\textit{Autoencoder}$. The name \textit{auto} signals that one part of the algorithm is the inverse of the other, so the input and output of an \textit{autoencoder} should be the same. In general, this correspondence is not perfect, as there is always some loss of information in a compression process, and the use of the word `inverse' is, in fact, an abuse of notation. However, a good \textit{autoencoder} should be able to recover the original image with high fidelity.

\begin{figure}[!ht]
    \centering
    \includegraphics[scale=0.35]{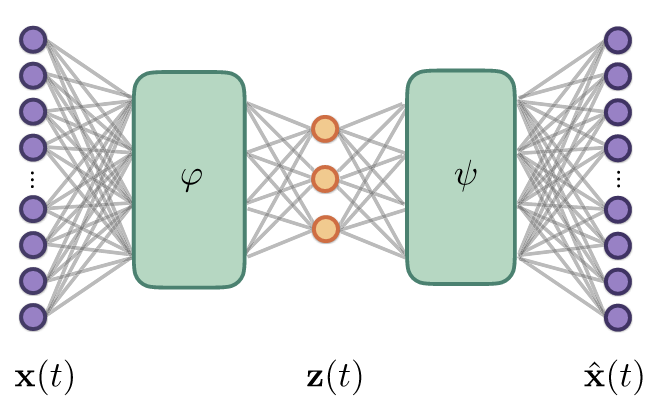}
    \caption{Pictorial representation of a neural network \textit{Autoencoder}. The input is represented by $\mathbf{x}(t)$, goes through a layer of neurons (usually CNN), and is then compressed into a smaller number of neurons by the \textit{encoder} $\varphi$, such that $\varphi(\mathbf{x(t)})=\mathbf{z}(t)$. The \textit{decoder}, the output, $\psi$ recovers the original input (approximately) with the operation $\psi(\mathbf{z}(t)) = \hat{\mathbf{x}}(t)$. The goal of the \textit{autoencoder} after training is to reproduce the relation $\psi\circ\varphi = \mathbb{I}$. Here, $\varphi$ is composed of CNN layers followed by fully connected layers, and $\psi$, the same, but with the order of the layers inverted. Source \cite{SINDyAutoencoder_steven2019} }
    \label{fig:autoencoder}
\end{figure} 

In the NN case, we can construct an architecture for an \textit{autoencoder}, as illustrated in Figure \ref{fig:autoencoder}. We have an input $\mathbf{x}(t)$ that provides information to an \textit{encoder} $\varphi$, whose output is a hidden layer with fewer neurons, $\mathbf{z}(t)$. In this example, and in the ones that follow, we assume that $\varphi$ is composed of a series of convolutional layers (CNN) followed by fully connected layers. Thus, similar to image compression, $\varphi$ reduces the input $\mathbf{x}(t)$, which contains a lot of information, to a few numbers in $\mathbf{z}(t)$, called the \textit{latent space} or \textit{latent layer} (sometimes also referred to as {\it code}). The task of the operator $\psi$ is then to decode the compressed information and recover the initial input approximately: $\psi(\mathbf{z}(t)) = \hat{\mathbf{x}}(t) \approx \mathbf{x}(t)$. It is common practice for the network of $\psi$ to be the reverse order of $\varphi$, i.e., fully connected layers followed by CNN layers. Therefore, the task of the \textit{autoencoder} is to approximate the relation $$\psi \circ \varphi \approx \mathbb{I},$$ where $\mathbb{I}$ is the identity operator in the vector space containing $\mathbf{x}(t)$, i.e., we want $\psi (\phi (\mathbf{x}(t))) \approx \mathbf{x}(t)$.

For didactic purposes, consider the following scenario where one can use an  \textit{autoencoder}: The reader intends to send an image to another person but is limited to transmitting only 3 real numbers. If we have a trained network, we can ``cut" the \textit{autoencoder} in half, keeping the \textit{encoder} and sending the \textit{decoder} to the recipient of the image. The original image is compressed by the \textit{encoder} $\varphi$, generating $\mathbf{z}(t)$. Thus, we can send $\mathbf{z}(t)$ (of size 3) to the other end of the communication channel, where it will be decoded with $\psi$, recovering the original message. 

In Physics, we have other interests. For example, in image recognition, these architectures can be used for cleaning original data. In observational techniques in astrophysics, \textit{autoencoders} can remove noise from electromagnetic waves, facilitating the visualization and processing of data \cite{Gheller_2021}.

\begin{figure}[!ht]
    \centering
    \includegraphics[width=1\linewidth]{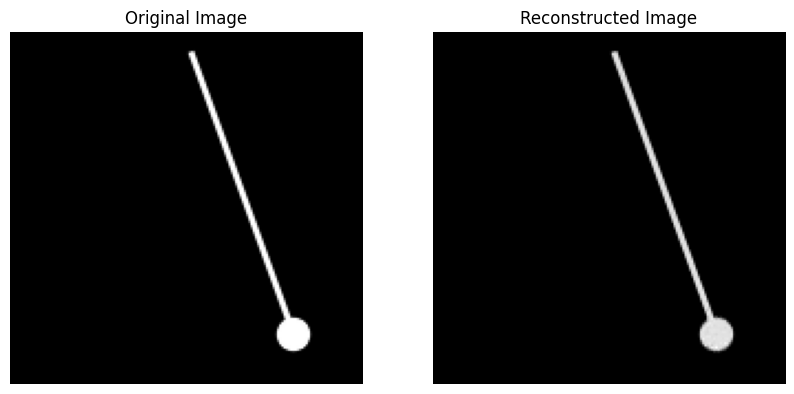}
    \caption{Comparison between the original image and the reconstructed image using an \textit{autoencoder} with a latent space of size 1. The image on the left represents the original image, while the image on the right is the reconstructed version of the same image.}
    \label{fig:imagem_reconstruida}
\end{figure}

In the following example, we will revisit the pendulum using \textit{autoencoders} and analyze the behavior of the latent space. Let us assume that our data consists of a sequence of \textit{frames} originating from a video of the dynamics of a pendulum with only one oscillation, starting from an initial angle $\theta_0=\theta(t=0)$ and zero initial velocity, $\dot{\theta}(0)=0$. To generate data analogous to a video of the pendulum's dynamics, our dataset is composed of a function that takes the numerical result of the pendulum's Equation (\ref{Eq:PenduloRaw}), $\theta(t)$, and converts it into a two-dimensional image $\mathbf{x}(t)$. 

The goal of this exercise is to verify whether the latent space of the \textit{autoencoder} with the pendulum images can be reduced to just one variable. This should be possible, given that the pendulum equation is a system that can be described by a single variable $\theta(t)$. We expect, therefore, that an \textit{autoencoder} with a latent space ($\mathbf{z}(t)$) of dimension one should be able to reproduce the original images, even in the nonlinear regime of large angles of the pendulum equation. Note that the nonlinearity of NN allows for the compression to occur even for the differential equation of large angles, which is not possible with linear dimensionality reduction techniques, such as \textit{Principal Component Analysis} (PCA) \cite[p.2]{SINDyAutoencoder_steven2019}, as these require linearization at some point during compression, and this can lead to losses in representation.

The result of our example can be seen in Figure \ref{fig:imagem_reconstruida}, where we compare the original image (left) and the reconstructed image (right) after training\footnote{In this example, it is recommended that the reader uses GPU processing (e.g., available for free on Google Colab), as the size of the network used requires long training phases.}.
\begin{figure}[!ht]
    \centering
    \includegraphics[width=1\linewidth]{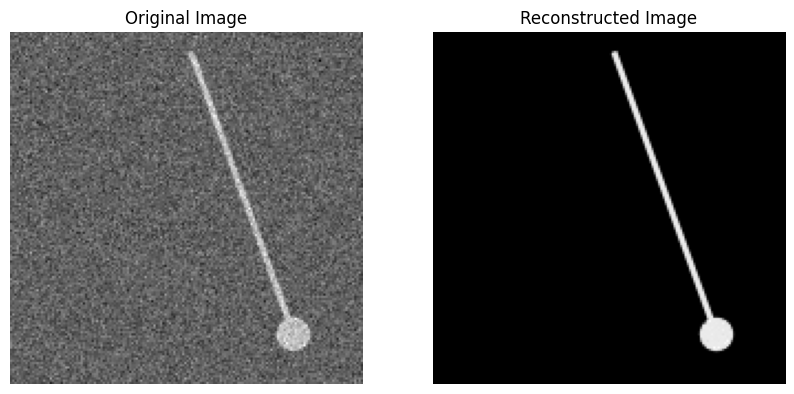}
    \caption{Comparison between the original image with added noise and the reconstructed image using an \textit{autoencoder} with a latent space of size 2. The image on the left represents the original image with added noise, while the image on the right is the reconstructed version of the same image. In this image, the noise level is $20\%$. In the supplementary material, we observed that the network still recovers the object and its position well even with noise levels of up to $40\%$.}
    \label{fig:imagem_reconstruida_wnoise}
\end{figure}
CNNs are well-known for abstracting only essential information from data, in the sense that their representations can ignore uncorrelated inputs in the data \cite{Vincent2008}. Thus, we can take a model trained only with images of the pendulum, without noise, and expose the model to noisy data (where we introduce uniformly distributed, hence uncorrelated, noise), as shown in Figure \ref{fig:imagem_reconstruida_wnoise}. Note that, despite the new images being perturbed by $20\%$ noise, i.e., with a probability $p=1/5$ of replacing a pixel with a random value, the model can still reproduce the expected behavior without the noise. Based on the trained models, it was found that by increasing the latent space dimension to 2, the noise could reach up to $40\%$ before the network showed signs of low performance.

This property is well-known in the field and is used, for example, in observational astrophysics for image cleaning \cite{Geller2022}. The tests discussed above can be found in the supplementary material, i.e., in the Notebook ``06-Example 3".

\begin{figure}[!ht]
    \centering
    \includegraphics[width=0.95\linewidth]{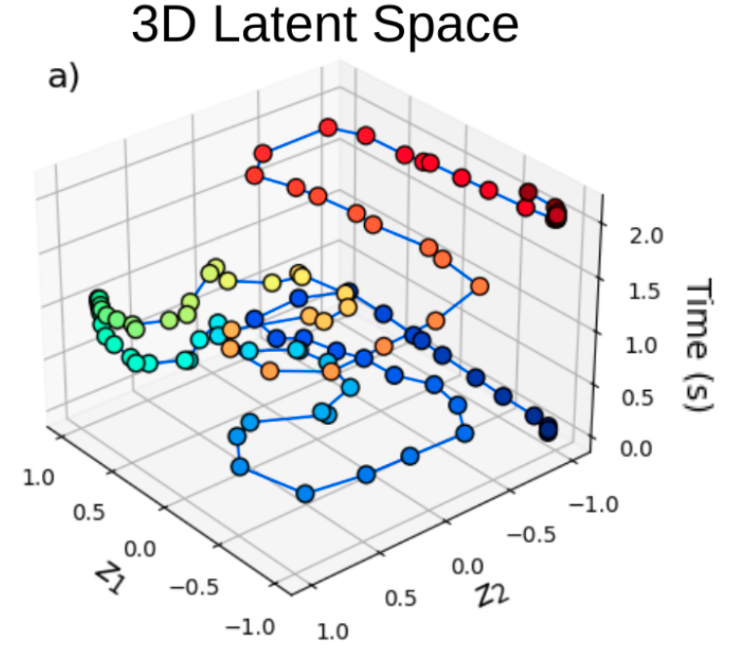}
    \includegraphics[width=1\linewidth]{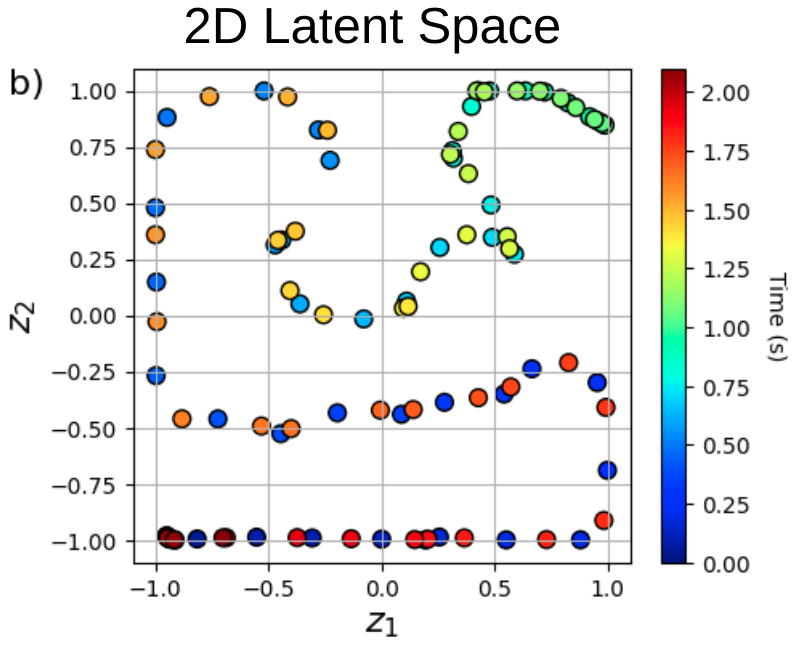}
    \caption{3D visualization in a) and 2D in b) of the latent space composed of $z_1$ and $z_2$, as a function of time. In both images, each point represents the position of the pendulum image in the latent space at a time instant $t_i$.}
    \label{fig:espaço-latente}
\end{figure}

There are still more possible use cases for a CNN, it can be used to analyze how each frame is distributed in the latent space, i.e., how each second is represented in the space encoded by the network. To illustrate this, we trained the network on a latent space of dimension two, $z_1(t)$ and $z_2(t)$, as shown in Figure \ref{fig:espaço-latente}. On the bottom image, we have a 2D plot, with the horizontal axis representing $z_1$ and the vertical axis $z_2$. Each circle represents the latent space for a frame of the video, corresponding to each time instant of the pendulum. To identify the time instant of the frame in each circle, we use a color, varying from 0 (blue) to $2.1$ seconds (red), as shown in the color map on the right side of the image. Note that along the curve there is no continuous separation of the pendulum's time instants in the figure on the right, i.e., according to the color map, we have blue and red colors interleaved, while these should ideally be at opposite ends of the continuous curve that generates these points: this is and effect of the non-linear nature of the encoding, resulting from the fact that, since the initial and final positions of the pendulum after a period $T$ should be very close, the encoding must map these data points close together, as they are almost the same image. However, although the network mapped the time intervals $[0,T/2]$ and $[T/2,T]$ in an overlapping manner, with opposite ordering, it is still possible to observe that the encoding is entirely contained within a single curve (i.e., a surface of dimension 1). To observe the continuity of the curve containing the encoding, we can add another axis, where we order the data in the same sequence as the frames are presented. The result can be seen in the top plot in Figure \ref{fig:espaço-latente} where we see the same plot in 3D. Thus, we can visualize the evolution of the latent space over time, and we see that the network correctly encoded the pendulum's position on a surface of dimension 1 parameterized by time.

This approach shows that we can leverage the representational capacity of autoencoders to estimate the dimensionality of a system. In such applications the complexity of the resulting encoding is of minimal interest, as one is only interested in the minimum dimension in which we can encode the system, as this indicates how many coordinates are needed to describe it.

\subsection{Example 4: Autoencoders and SINDy Architectures for ODE Discovery.}
\label{sec:EX:SINDY}
As discussed in the introduction, there are sensible reasons to adopt a skeptical stance toward NNs, as their functioning can sometimes be more opaque than the system that are being studied. After all, a NN can have thousands (or even trillions, in recent cases) of parameters to be adjusted, which act non-linearly, surpassing human analytical capacity \cite[p.4]{MSchuld2021}. That said, there is an effort in the community to combine prior knowledge about physical systems to reduce the opacity of NN and obtain results that provide a more detailed perspective on how a NN operates, without loss of generality.

In this final example, we will apply the concepts developed in the previous examples to explore a recent architecture of particular interest to the scientific community. The focus will be on Sparse Identification of Nonlinear Dynamics (SINDy) \cite{SINDyAutoencoder_steven2019}. The SINDy method aims to identify the underlying dynamics of a nonlinear physical system by constraining the latent space of the NN in a way that reveals its equations of motion, while the network, as a whole, maintains the functionality of an autoencoder.

Suppose that we have a nonlinear system, in this case generated by Equation (\ref{Eq:PenduloRaw}), which produces a dataset of images $\mathbf{x}(t)$. For example, a video of a pendulum with large angles. Suppose also that the latent space can be represented by a differential equation (possibly nonlinear), such as
\begin{equation} \label{eq:latent_space_eq}
    \frac{d}{dt}\mathbf{z}(t) = \mathbf{g}(\mathbf{z}(t)),
\end{equation} 

\noindent or as
\begin{equation} \label{eq:latent_space_eq_2nd}
    \frac{d^2}{dt^2}\mathbf{z}(t) = \mathbf{g}(\mathbf{z}(t)),
\end{equation} 
\noindent in a second degree ODE case.

We shall use the first derivative of $\boldsymbol{z}$ as the canonic use of SINDys in the following explanation. However, we will be interweaving the explanation with examples of the second derivative case, since this is the model of the non-linear pendulum system. In principle, we could have chosen any derivative order, as long as it represents the model of a physical system. Although we will be presenting only the first and second derivative case (in parallel), it should be clear that the same rules apply for higher derivative cases. The example for the second derivative case for the pendulum can be explored in more details in the Jupyter notebooks.

If our \textit{autoencoder} is capable of recovering the original images with a single variable in $\mathbf{z}(t)$ (in the latent space), which we can assume to be an affine transformation of $\theta(t)$ (the function generating the image dataset, i.e., that is also the solution to equation (\ref{Eq:PenduloRaw})), then we can require that it also respects the same differential equation, i.e., that it is a solution to 

\begin{equation}
    \ddot{z}(t) + \sin{(z(t))} =0.
\end{equation} 

\noindent Remember that $z$ and $\theta$ are functions defined in different spaces.

But suppose for now that we are unaware of the pendulum equation and we only have the raw data in our hands, i.e., our dataset is a set of images of the type $\{\mathbf{x}(t), \dot{\mathbf{x}}(t), \ddot{\mathbf{x}}(t)\}$ (remember that we can take the derivative of an image numerically, without knowing its analytical expression, as long as $\mathbf{x}(t)$ is sufficiently smooth). We will address the following question in what follows: how can we ensure that the equation for $z$ truly represents our system?

To do this, we create a library of $m$ possible functions for the system $\{z, z^2, \ldots, z^n, \sin(z), \cos(z), \sqrt{z}, 1/z, \ldots\}$, each of which is an entry in a matrix of functions $\boldsymbol{\Theta}$ (here, of size $1 \times m$). We will also define a matrix ($m \times 1$) of coefficients $\boldsymbol{\Xi} = (\xi_1, \ldots, \xi_m)$, so that Equation (\ref{eq:latent_space_eq}) can be rewritten as:
\begin{equation} \label{eq:latent_space_SINDy}
    \dot{z}(t) = \boldsymbol{\Theta}(z(t))\boldsymbol{\Xi}.
\end{equation}

\noindent In the second order case, instead of equation (\ref{eq:latent_space_SINDy}), we take \begin{equation} \label{eq:latent_space_SINDy_2nd}
    \ddot{z}(t) = \boldsymbol{\Theta}(z(t))\boldsymbol{\Xi}.
\end{equation}
Note that equation (\ref{eq:latent_space_SINDy_2nd}) is not the second derivative of equation (\ref{eq:latent_space_SINDy}). It is an alternative description to be used when we are examining systems that are characterized by second order ODEs.

\begin{figure} [h]
    \centering
    \includegraphics[scale=0.36]{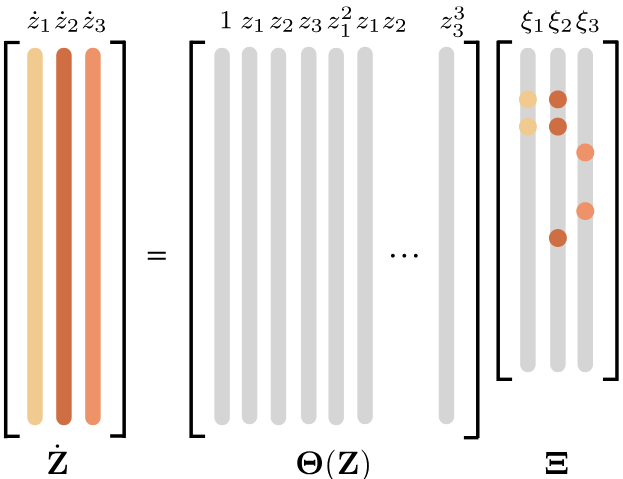}
    \caption{Generalization of Equation (\ref{eq:latent_space_SINDy}) for a multivariable system. The SINDy autoencoder performs a search among various functions in a library $\boldsymbol{\Theta}$, where the derivative of the latent space $\mathbf{z}$ is a linear combination of the library with coefficients given by $\boldsymbol{\Xi}$. This equation is used as part of the cost function for optimization, which iteratively suppresses some coefficients $\xi_{ij}$ until $\boldsymbol{\Xi}$ is sufficiently sparse, and the equation returns a simple model. A successful optimization ensures that the latent space provides an analytical expression for the equations of motion while containing few terms, as a consequence of the sparsity of the found model. Source \cite{SINDyAutoencoder_steven2019}.}
    \label{fig:SINDy_latent_space_Model}
\end{figure}

To calculate the derivative of $\dot{z}$ with respect to time, we use the expression $\dot{\mathbf{x}}(t)$ through the chain rule, in the same way we did for the \textit{backpropagation} algorithm in Section \ref{sec:RNProfunda}. In this case, using the notation from Figure \ref{fig:autoencoder}, we have that $z = \varphi(\mathbf{x})$, so
\begin{equation}
    \dot{z} = \nabla_x \varphi (\mathbf{x}) \dot{\mathbf{x}}.
\end{equation}

\noindent So now we have an expression for $\dot{z}$ in terms of $\dot{\mathbf{x}}$. The second derivative is taken in the same way; see the supplementary material in \cite{SINDyAutoencoder_steven2019}, where we obtain
\begin{equation} \label{eq:SINDy_2ndDeriv}
    \ddot{z} = \nabla_x^2 \varphi (\mathbf{x}) \dot{\mathbf{x}}^2 + \nabla_x \varphi (\mathbf{x}) \ddot{\mathbf{x}}.
\end{equation}

Similarly, we can impose that the derivative of the newly generated image, i.e., the outputs of the NN  $\dot{\tilde{\mathbf{x}}}(t)$, faithfully reproduces the derivatives of the images in the dataset, $\dot{\mathbf{x}}(t)$. This constraint, applied through the minimization of a cost function defined below, ensures that $z(t)$ is a representative model of the system, just like the original model in $t$. In our case, we will impose this condition on the second derivative, i.e., $\ddot{\tilde{\mathbf{x}}} \approx \ddot{\mathbf{x}}$.

Furthermore, for $z$ to also be a model of the generated images, we will require that the relation $$\frac{d^n}{dz^n}\tilde{\mathbf{x}}(z) \approx \frac{d^n}{dt^n}\mathbf{x}(t),$$ holds, and thus that $z$ behaves like $\theta$, not only for the dataset but also for the generated data. Therefore, still using the notation from Figure \ref{fig:autoencoder}, where $\psi(z) \approx \tilde{\mathbf{x}}$, we have that
\begin{equation} \label{eq:Sindy_derivativeApprox}
    \dot{\mathbf{x}} \approx \frac{d}{dz}\tilde{\mathbf{x}} = \nabla_z \psi (z) \boldsymbol{\Theta}(z) \boldsymbol{\Xi}.
\end{equation}

\noindent Where we have substituted $\dot{z}$ by its expression in equation (\ref{eq:latent_space_SINDy}). In the case of a second order equation, we should have instead
\begin{equation} \label{eq:Sindy_derivativeApprox}
    \ddot{\mathbf{x}} \approx \frac{d^2}{dz^2}\tilde{\mathbf{x}} = \nabla^2_z \psi (z) \dot{z}^2 +  \nabla_z \psi (z)\boldsymbol{\Theta}(z) \boldsymbol{\Xi}.
\end{equation}
\noindent Were now we substitute $\ddot{z}$ by equation (\ref{eq:latent_space_SINDy_2nd}).

In this way, for the case of the first-order derivative, we have three requirements for our network, which translates into three terms for our cost function: First, the same term we had in Example \ref{sec:EX:AutoencoderPendulo}, for the reconstruction of the original image,
\begin{equation} \label{eq:SINDy_reconloss}
    \mathcal{L}_{\text{recon}} = || \mathbf{x} - \psi \circ \varphi (\mathbf{x})||^2_2,
\end{equation}

\noindent added to the second term, for the reconstruction of the derivative of the output image in terms of $z$
\begin{equation}
    \mathcal{L}_{\dot{\mathbf{x}}} = ||\dot{\mathbf{x}} - \nabla_z \psi (z) \boldsymbol{\Theta}(z) \boldsymbol{\Xi} ||^2_2,
\end{equation}

\noindent and finally the term related to the derivative of the model generated by $z$
\begin{equation}
    \label{eq:SINDy_dzloss}
    \mathcal{L}_{\dot{z}} =  || \nabla_x \varphi(\mathbf{x})\dot{x} -  \boldsymbol{\Theta}(z(t))\boldsymbol{\Xi} ||^2_2.
\end{equation}

\noindent In the case of models for systemsof second order ODE, we must redeclare the two partial cost functions above as 
\begin{equation}
    \mathcal{L}_{\ddot{\mathbf{x}}} = ||\ddot{\mathbf{x}} - (\nabla^2_z \psi (z) \dot{z}^2 +  \nabla_z \psi (z)\boldsymbol{\Theta}(z) \boldsymbol{\Xi}) ||^2_2,
\end{equation}
\noindent and the term of the derivative of the model generated by $z$
\begin{equation}
\label{eq:SINDy_dzloss_2nd}
    \mathcal{L}_{\ddot{z}} =  ||\nabla_x^2 \varphi (\mathbf{x}) \dot{\mathbf{x}}^2 + \nabla_x \varphi (\mathbf{x}) \ddot{\mathbf{x}} - \boldsymbol{\Theta}(z) \boldsymbol{\Xi}||^2_2.
\end{equation}

It is important to note that the calculation of these functions is efficient in the following sense: just as for \textit{backpropagation}, we can store the values of the derivatives $\nabla \varphi$ and $\nabla \psi$ when calculating $\tilde{\mathbf{x}}$, and use both for the calculation of $\mathcal{L}$ and for updating the weights and biases. With the terms defined above, we have the necessary tools to find analytical models for the dynamics in terms of $z$. 

However, we have not yet explained how the model can be sparse, as suggested by the title of the architecture. Note that if we optimize the network as it stands, we will obtain models of the type $\ddot{z} = \xi_1 z + \xi_2 z^2 \ldots \xi_{n+1} \sin(z) + \ldots$. This is not a particularly informative model about the system, as it has as many terms as given. Two new constraints are then necessary: first, we will require that $\boldsymbol{\Xi}$ has a small $||.||_1$ norm, forcing the values of $\xi_i$ to be close to the interval $(-1,1)$. Additionally, after a few optimization steps, the values of $\xi_i$ will be distributed such that some will be larger than others (in absolute value), simply because they are more relevant to the dynamics than others. Thus, we add a cutoff in the optimization that eliminates coefficients $|\xi_i| < 0.1$, setting $\xi_i=0$ after the cutoff. With this method, we are eliminating elements from the matrix $\boldsymbol{\Xi}$ and making it sparse, forcing our model to be as simple as possible.

Our cost function then takes the final form:
\begin{equation} \label{eq:SINDy_lossfunction}
    \mathcal{L}_{\text{SINDy}} = \mathcal{L}_{\text{recon}} + \lambda_1\mathcal{L}_{\dot{x}} + \lambda_2\mathcal{L}_{\dot{z}} + \lambda_3||\boldsymbol{\Xi}||_1,
\end{equation}
\noindent where $\lambda_i$ are hyperparameters to be adjusted, and for the pendulum case, the values used were $ {\lambda_1 =5 \cross 10^{-4}, \lambda_2 =5 \cross 10^{-5}, \lambda_3 = 10^{-5}}$.

\begin{figure} [ht]
    \centering
    \includegraphics[scale=0.45]{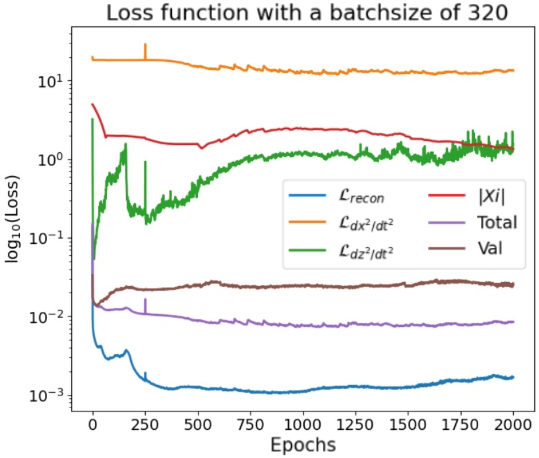}
    \caption{Multiple loss functions of the SINDy autoencoder model as a function of training epochs, each defined by equations (\ref{eq:SINDy_reconloss}–\ref{eq:SINDy_dzloss}). The y-axis shows the error metric on a logarithmic scale, and the x-axis shows the number of epochs. The curve corresponding to the total loss function (Total, purple), given by equation (\ref{eq:SINDy_lossfunction}), shows a decreasing trend, reflecting the global optimization of the model. A peak at 250 epochs represents the stage where small coefficients $|\xi_i| < 0.1$ are pruned to enforce sparsity in the model. At this point, the network’s performance is temporarily degraded. The brown curve (Val) represents the loss function computed on the validation data—data never seen by the network—to evaluate its performance outside the training set.}
    \label{fig:SINDy_loss}
\end{figure}
Note that there is a fundamental difference between two possible approaches to training a network. In the present case, the network is trained from the beginning with a cost function that imposes the need to find and respect a differential equation while also optimizing the image generation simultaneously. In a second approach, the network is first trained for image recognition, and then the latent space is adjusted independently. In the second case, by optimizing the NN in two distinct stages, each stage operates in different parameter space surfaces. Consequently, the first optimization may lead to a local minimum far from the optimal solutions for the second. 
In the case of SINDy architectures, since both conditions are optimized simultaneously, the NN seeks a path in the parameter space that satisfies all constraints at the same time.
The priority assigned to each parameter is, to some extent, determined by the hyperparameters $\lambda_i$. As with other hyperparameters, the appropriate choice of these values requires familiarity and experience with the technique \cite[p. 6]{carleo_machine_2019}.

The SINDy architecture can be used for nonlinear systems with more variables, such as Lorentz attractors and diffusion reactions \cite{SINDyAutoencoder_steven2019}. In this case, Equation (\ref{eq:latent_space_SINDy}) becomes a matrix system as in Figure \ref{fig:SINDy_latent_space_Model}. In this case, we also need to include cross terms like ${\{z_1^{k_1}z_2^{k_2}\ldots z_q^{k_q} | k_1,k_2,\ldots, k_q \in \{1,2,3, \ldots, n\}\}}$, where $k_q$ is the order of the monomial corresponding to dimension $q$, in the library $\boldsymbol{\Theta}$, to reproduce systems of coupled equations.

In the supplementary material of this article, inspired by previous work in \cite{GithubSINDy}, we have the worked example for the nonlinear pendulum equation, where our dataset consists of images of the type $\{\mathbf{x}(t), \dot{\mathbf{x}}(t), \ddot{\mathbf{x}}(t)\}$. Here, we constrain the pendulum's energy to the case where ${|\dot{\theta}^2(0)/2 - \cos{(\theta(0))}|\leq 0.99}$, to avoid complete (or double) rotations in the oscillation. For validation of the training, we compare the three $\{\mathbf{x}, \dot{\mathbf{x}}, \ddot{\mathbf{x}}\}$ with $\{ \tilde{\mathbf{x}}, \tilde{\dot{\mathbf{x}}}, \tilde{\ddot{\mathbf{x}}}\}$, and although we calculate the difference in states and accelerations in the cost function, the network is still capable of recovering the velocity at times, since the information of $\dot{\mathbf{x}}$ is implicit in Equation (\ref{eq:SINDy_2ndDeriv}). Thus, even though it is not explicitly part of the training to take the difference with the first derivative, we recover an approximate behavior, demonstrating that the learning is robust.
\begin{figure} [h]
        \centering
        \includegraphics[width=1\linewidth]{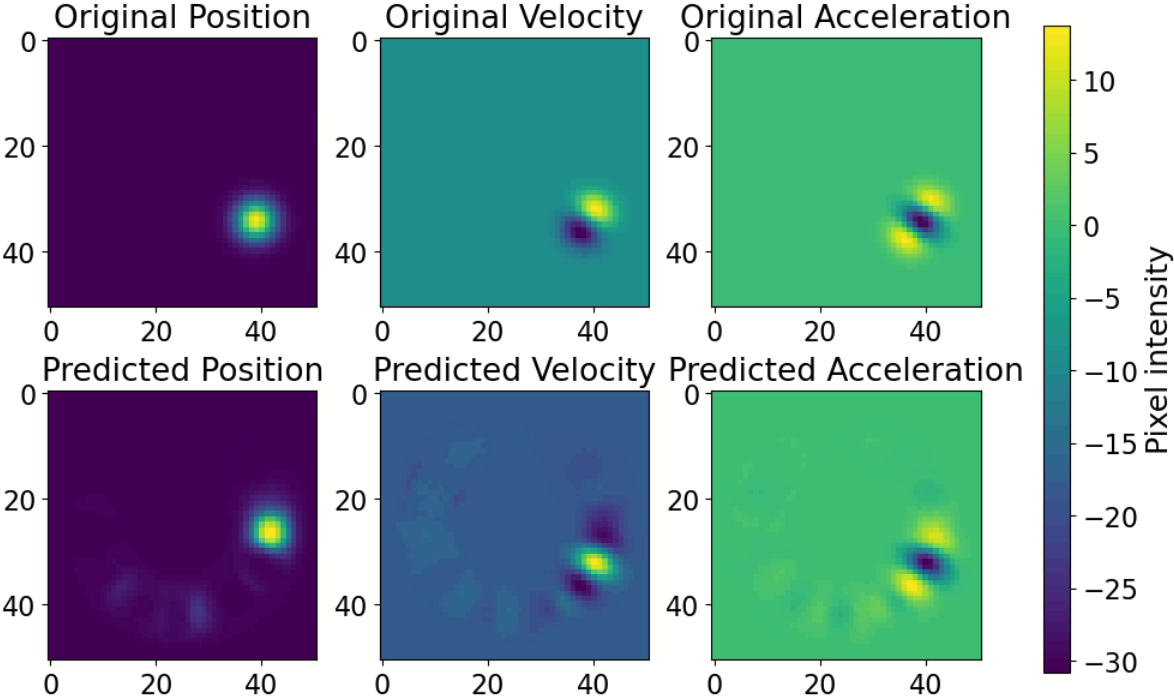}
        \caption{Jupyter Notebook image comparing the input and output of SINDy for validation data of position, velocity, and acceleration, i.e., comparing $\{\mathbf{x}, \dot{\mathbf{x}}, \ddot{\mathbf{x}}\}$ with $\{ \tilde{\mathbf{x}}, \tilde{\dot{\mathbf{x}}}, \tilde{\ddot{\mathbf{x}}}\}$ after 2000 epochs. Note that in this optimization, velocity data was not used explicitly during training (only implicitly, through equation (\ref{eq:SINDy_2ndDeriv})), yet the model is still able to approximately recover it.}
        \label{fig:SINDy_ValidationData}
\end{figure}

For our example, we used the polynomial library up to degree 4 and the sine function. Among the 10 initializations we tested, 8 recovered the correct model, i.e., equation (\ref{Eq:PenduloRaw}), while the other 2 did not recover any model. It is worth noting that in some initializations the network recovered the model from equation (\ref{Eq:PenduloSmallAngle}), which is the linear approximation of the original equation, while at other times it recovered equations of the form $\ddot{z} = \xi_0\sin{z} + \xi_1 z$, eliminating higher polynomial terms and other functions as zero, ensuring that the solution remains sparse.

The results of the generated images $\{ \tilde{\mathbf{x}}, \tilde{\dot{\mathbf{x}}}, \tilde{\ddot{\mathbf{x}}}\}$ are shown in Figure \ref{fig:SINDy_ValidationData}, compared with the validation data $\{\mathbf{x}, \dot{\mathbf{x}}, \ddot{\mathbf{x}}\}$ in a model that found the pendulum equation. Notice that we optimized the model only for position and acceleration, yet it was still able to approximate the result for the velocity images, demonstrating that the network indeed learned the model and is capable of reproducing lower-order derivatives.

Furthermore, these networks (and PINNs in general) require less training data than traditional autoencoders. It is noteworthy that in this example we did not use CNNs, only a fully connected network, and even so, fewer epochs were needed than in other cases to achieve satisfactory results, we found that in general around $15\%$of the data was required in other architectures, this means that if 2000 epochs were needed in another architecture, the SINDy should be able to find the right model in a range of 300-500 epochs, as can be seen by the sharp decline in the first epochs in Figure \ref{fig:SINDy_loss}. This case serves as an example of how analytical knowledge reduces the need for training through examples, i.e., once the network has learned the model through optimization, it requires less information to reproduce the data than models discussed in the rest of this article.

\section{Conclusion}\label{sec8}

In this paper, an introduction to basic and advanced aspects of NN was presented, from the perceptron to more complex applications in Physics. We reviewed basic concepts and demonstrated through examples how NNs can be employed in binary classification tasks and linear regression. As for the Applied Physics examples, we have shown the effectiveness of deep NNs applied to a single physical system (the pendulum), exploring the same problem under various approaches in increasing difficulty, such as: how to find the physical parameter of a function from data, determine the solution to a differential equation, and identify which differential equation describes a given set of data.

Among these examples, we highlight the use of PINNs as a particularly promising numerical tool, although these were used only to solve differential equations, they demonstrate a range of possibilities still unexplored. The PINNs approach is particularly relevant for physicists, as it allows NNs to learn more efficiently by integrating physical constraints in the learning phase, reducing the need for large datasets, and improving the accuracy of predictions for systems governed by previously known laws. Finally, we observed that through the use of SINDy architectures, NNs are capable of learning physical systems analytical models even in nonlinear systems, potentially helping not only in studies of systems that lack such models but also providing greater clarity on the output of the network's learning, as it is expressed as an explicit equation.

This study highlights the growing importance of NNs as powerful tools in Physics research, demonstrating their potential to revolutionize traditional approaches in Theoretical and Computational Physics. By integrating advanced ML methods, such as NNs, into physical modeling, this work not only demonstrates the accuracy of simulations and the analysis of complex data but also promotes a shift in the way science is conducted. This interdisciplinary integration allows for the exploration of new horizons in problems that were previously intractable or required significant approximations. Thus, it paves the way for a promising new scientific paradigm, where artificial intelligence techniques and physical methods complement each other to advance the understanding of natural phenomena.

\section*{Supplementary Material}
The Jupyter notebooks associated with the results worked on in the examples can be found in \cite{Github}.
 \section*{Acknowledgments}
We would like to thank Prof. Dr. Celso Jorge Villas Boas for his suggestions for improving the work.

This work was supported by the National Council for Scientific and Technological Development (CNPq), Process No. 465469/2014-0, and the São Paulo Research Foundation (FAPESP), Process No. 2023/15739-3.


\section*{Appendix A - Calculation of the cost function}
\label{sec:Apendix:CalcFuncCusto}

\begin{table}[ht]
    \centering
    \renewcommand{\arraystretch}{2.5} 
    \begin{tabular}{|c|c|c|}
    \hline
    \textbf{Activation Function} & $f(x) $ & \textbf{Derivative} \\
    \hline
    Sigmoid & $ \frac{1}{1 + e^{-x}}$ & $f'(x) = f(x)(1 - f(x))$ \\
    \hline
    Tanh & $ \tanh(x)$ & $f'(x) = 1 - f(x)^2$ \\
    \hline
    ReLU & $ \max(0, x)$ & $f'(x) = \begin{cases} 1 & \text{if } x > 0 \\ 0 & \text{if } x \leq 0 \end{cases}$ \\
    \hline
    Leaky ReLU & $ \max(0.01x, x)$ & $f'(x) = \begin{cases} 1 & \text{if } x > 0 \\ 0.01 & \text{if } x \leq 0 \end{cases}$ \\
    \hline
    \end{tabular}
    \caption{Comparison of usual activation functions and their derivatives}
    \label{tab:activation_functions}
\end{table}

Each ML task (such as regression, classification, etc) has a cost function that is typically better suited for it. On Table \ref{tab:activation_functions} there are some of the main cost functions used in ML.

For illustrative purposes, we present the graphs of the activation functions and their derivatives as well.

\begin{figure}[!ht]
    \centering
    \includegraphics[width=0.8\linewidth]{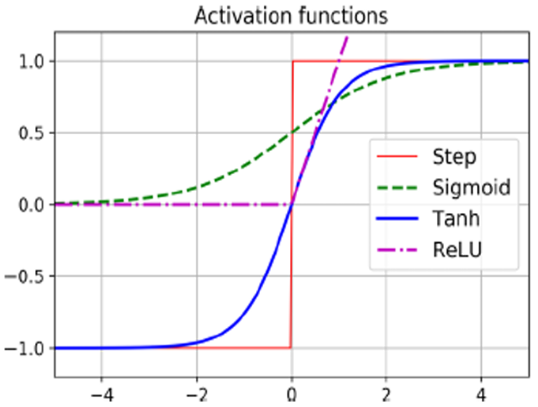}
    \includegraphics[width=0.8\linewidth]{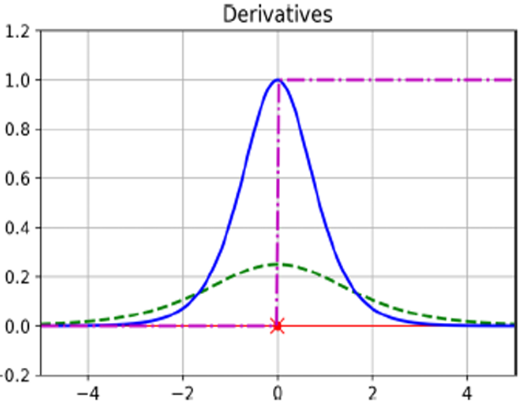}
    \caption{Activation functions (above) cited in Table \ref{tab:activation_functions}, and their derivatives (below). Source: \cite{geron}.}
\end{figure}

Notice that for large values some functions derivatives tend to zero, which affects the training process. Table II presents some of the cost functions, such as Mean Squared Error (MSE), one of the most common cost functions for regression tasks, it measures the mean of the squares of the differences between the predicted and real values. Mean Absolute Error (MAE): Also used in regression, MAE measures the mean of the absolute differences between the predicted and real values, provides an error measure that is robust to outliers. Cross-Entropy Loss (CCE) is widely used in classification problems. This cost function measures the difference between two probability distributions, the one predicted by the model and the true distribution. For binary classification, it is known as Binary Cross-Entropy (BCE). Note: BCE Loss is a specific case of CCE, where we have only two classes.

During the calculation of the weight update, we need the gradient of the cost function with respect to the parameters. So we will present an example of the derivation of the expression using as examples the MSE cost function and the sigmoid activation function. 

\begin{table} [h]
    \centering
    \renewcommand{\arraystretch}{2.5} 
    \begin{tabular}{|c|c|c|}
    \hline
    \textbf{Loss} & \textbf{Mathematical Expression}  \\ \hline
    MSE & $\frac{1}{n} \sum_{i=1}^n (y_i - \hat{y}_i)^2$  \\ \hline
    MAE & $\frac{1}{n} \sum_{i=1}^n |y_i - \hat{y}_i|$  \\ \hline
    BCE & $-\sum [y \log(\hat{y}) + (1-y) \log(1-\hat{y})]$ \\ \hline
    CCE & $-\sum_{i=1}^n \sum_{j=1}^m y_{ij} \log(\hat{y}_{ij})$ \\\hline
    \end{tabular}
    \caption{Examples of Activation Functions and Cost Functions.}
\end{table}

Take the gradient of $\mathcal{L}$ with respect to $w$,

\begin{align*}
    \nabla_w \mathcal{L} &=\frac{d}{dw} (y - \hat{y})^2, \\
    \nabla_w \mathcal{L} &= -2(y - \hat{y}) \frac{d}{dw}\hat{y}, \\
    \nabla_w \mathcal{L} &= -2(y - \hat{y}) \frac{d}{dw}f(w^T \mathbf{x}), \\
    \nabla_w \mathcal{L} &= -2(y - \hat{y}) f' \frac{d}{dw}(w^T \mathbf{x}), \\
    \nabla_w \mathcal{L} &= -2(y - \hat{y}) f' \mathbf{x} \frac{d}{dw}w^T, \\
    \nabla_w \mathcal{L} &= -2(y - \hat{y}) f' \mathbf{x}. 
\end{align*}

Given the expression, for $w$ we can do the same for $b$.
\begin{align*}
    \nabla_b \mathcal{L} &= \frac{d}{db} (y - \hat{y})^2, \\
    \nabla_b \mathcal{L} &= -2(y - \hat{y}) \frac{d}{db}\hat{y}, \\
    \nabla_b \mathcal{L} &= -2(y - \hat{y}) \frac{d}{db}f(w^T \mathbf{x}), \\
    \nabla_b \mathcal{L} &= -2(y - \hat{y}) f' \frac{d}{db}(w^T \mathbf{x}), \\
    \nabla_b \mathcal{L} &= -2(y - \hat{y}) f' \mathbf{x} \frac{d}{db}w^T, \\
    \nabla_b \mathcal{L} &= -2(y - \hat{y}) f'. 
\end{align*}

Thus,
\begin{align*}
    \nabla_w \mathcal{L} &= -2(y - \hat{y}) f' \mathbf{x},  \\
    \nabla_b \mathcal{L} &= -2(y - \hat{y}) f' ,
\end{align*}

\noindent where \( f' \) depends on the activation function being used. In Table \ref{tab:activation_functions}, we have some examples; for illustration, we will use the Sigmoid \( f' = f(1 - f) \), where \( f = \hat{y} \).
\begin{align*}
    \nabla_w \mathcal{L} &= -2(y - \hat{y}) \hat{y}(1- \hat{y}) \mathbf{x},  \\
    \nabla_b \mathcal{L} &= -2(y - \hat{y}) \hat{y}(1- \hat{y}).
\end{align*}



\begin{thebibliography}{9}

\bibitem{geron}
Géron, A. (2017). Hands-on Machine Learning with Scikit-Learn, Keras, and TensorFlow. O'Reilly Media. ISBN: 978-1-4920-3264-9.

\bibitem{NathanKutz2017}
Brunton, S. L. and Kutz, J. N. Data Driven Science and Engineering. (2017).

\bibitem{carleo_machine_2019}
Carleo, Giuseppe et al, Machine learning and the physical sciences. Reviews of Modern Physics. \textbf{91}, 045002 (2019).

\bibitem{Schmidhuber2015}
Schmidhuber, J. (2015). Deep learning in neural networks: An overview. Neural Networks, 61, 85-117. ISSN 0893-6080. 

\bibitem{KNN}
Uddin, S., Haque, I., Lu, H. et al. Comparative performance analysis of K-nearest neighbour (KNN) algorithm and its different variants for disease prediction. Sci Rep 12, 6256 (2022).

\bibitem{KNN2}
Cunningham, Padraig, and Sarah Jane Delany. "k-Nearest neighbour classifiers: (with Python examples)." arXiv preprint arXiv:2004.04523 (2020).


\bibitem{logiste-regression1}
Peng, Chao-Ying Joanne, et al. “An Introduction to Logistic Regression Analysis and Reporting.” The Journal of Educational Research, vol. 96, no. 1, pp. 3–14,(2002).

\bibitem{logiste-regression2}
Bertsimas, D.,  King, A. Logistic Regression: From Art to Science. Statistical Science, 32, 367-384,  (2017).

\bibitem{SVM1}
Noble, W. What is a support vector machine?. Nat Biotechnol 24, 1565–1567 (2006).

\bibitem{SVM2}
Lorena, A. C., de Carvalho, A. C. P. L. F. Uma Introdução às Support Vector Machines. Revista De Informática Teórica E Aplicada, 14(2), 43–67,(2007).

\bibitem{decision-tree1}
Kotsiantis, S.B. Decision trees: a recent overview. Artif Intell Rev 39, 261–283 (2013).

\bibitem{decision-tree2}
Kingsford C, Salzberg SL. What are decision trees? Nat Biotechnol. Sep;26(9):1011-3 (2008). 

\bibitem{supervised-nn}
Neupert, Titus, et al. "Introduction to machine learning for the sciences." arXiv preprint arXiv:2102.04883 (2021).

\bibitem{unsupervised}
Giuseppe Bonaccorso. Machine Learning Algorithms: A reference guide to popular algorithms for data science and machine learning. Packt Publishing, (2017)

\bibitem{Marquardt2021}
Marquardt, F. Machine learning and quantum devices. SciPost Physics Lecture Notes (2021)

\bibitem{Goodfellow-et-al-2016}
Goodfellow, I., Bengio, Y. and Courville, A. Deep Learning. MIT Press (2016). Available at: http://www.deeplearningbook.org.

\bibitem{SINDyAutoencoder_steven2019}
Champion, K., Lusch, B., Kutz, J. N. and Brunton, S. L. Data-driven discovery of coordinates and governing equations. Proc. Natl. Acad. Sci. U.S.A. 116, 22445–22451 (2019).

\bibitem{alphago}
Silver, D., Huang, A., Maddison, C. et al. Mastering the game of Go with deep neural networks and tree search. Nature 529, 484–489 (2016). https://doi.org/10.1038/nature16961

\bibitem{atari}
Mnih, V., Kavukcuoglu, K., Silver, D. et al. Human-level control through deep reinforcement learning. Nature 518, 529–533 (2015). https://doi.org/10.1038/nature14236

\bibitem{robothand}
OpenAI, et al. Solving Rubik’s Cube with a Robot Hand. arXiv:1910.07113, 2019. https://doi.org/10.48550/arXiv.1910.07113.

\bibitem{Gligorov2013}
Gligorov,  V V and Williams,  M, Efficient,  reliable and fast high-level triggering using a bonsai boosted decision tree, Journal of Instrumentation. \textbf{02}, vol 8 

\bibitem{baldi2014}
Baldi, P., Sadowski, P. and Whiteson, D. Searching for exotic particles in high-energy physics with deep learning. Nat Commun 5, 4308 (2014).

\bibitem{Feickert}
Feickert, Matthew, and Benjamin Nachman. "A living review of machine learning for particle physics." arXiv preprint arXiv:2102.02770 (2021).

\bibitem{carrasquila2017}
Carrasquilla, J., Melko, R. Machine learning phases of matter. Nature Phys 13, 431–434 (2017).

\bibitem{VonNieuwenburg2017}
Van Nieuwenburg, E., Liu, YH. and Huber, S. Learning phase transitions by confusion. Nature Phys 13, 435–439 (2017). 

\bibitem{Smith2017}
Smith, J. S., Isayev, O., Roitberg, A. E. (2017). ANI-1: an extensible neural network potential with DFT accuracy at force field computational cost. Chemical Science, 8(4), 3192-3203. The Royal Society of Chemistry.

\bibitem{PIML-review}
Hao, Zhongkai, et al. "Physics-informed machine learning: A survey on problems, methods and applications." arXiv preprint arXiv:2211.08064 (2022).

\bibitem{PIML-review-nature}
G. E. Karniadakis, I. G. Kevrekidis, L. Lu, P. Perdikaris, S. Wang, and L. Yang, “Physics-Informed machine learning,” Nature Reviews Physics, vol. 3, no. 6, pp. 422–440, 2021.

\bibitem{PIML-review2}
Meng, Chuizheng, et al. "When physics meets machine learning: A survey of physics-informed machine learning." arXiv preprint arXiv:2203.16797 (2022).


\bibitem{PINN-review}
Cuomo, Salvatore, et al. "Scientific machine learning through physics–informed neural networks: Where we are and what’s next." Journal of Scientific Computing 92.3 (2022): 88.

\bibitem{PINN-review2}
Faroughi, Salah A., et al. "Physics-guided, physics-informed, and physics-encoded neural networks in scientific computing." arXiv preprint arXiv:2211.07377 (2022).


\bibitem{PINN-review-fluid}
S. Cai, Z. Mao, Z. Wang, M. Yin, and G. E. Karniadakis, “Physics informed neural networks (pinns) for fluid mechanics: A review,”  arXiv preprint arXiv:2105.09506, 2021

\bibitem{PINN-review-fluid2}
Sharma, P.; Chung, W.T.; Akoush, B.; Ihme, M. A Review of Physics-Informed Machine Learning in Fluid Mechanics. Energies 2023, 16, 2343.

\bibitem{Psaros}
A. F. Psaros, X. Meng, Z. Zou, L. Guo, and G. E. Karniadakis, “Uncertainty quantification in scientific machine learning: Methods, metrics, and comparisons,” arXiv preprint arXiv:2201.07766,
 2022.

\bibitem{WangandR}
 R.WangandR.Yu,“Physics-guided deep learning for dynamical systems: A survey,” arXiv preprint arXiv:2107.01272, 2021.

\bibitem{Carrasquilla}
Carrasquilla, Juan, and Giacomo Torlai. "Neural networks in quantum many-body physics: a hands-on tutorial." arXiv preprint arXiv:2101.11099 (2021).

\bibitem{nn-phonics}
Pedro Freire, Egor Manuylovich, Jaroslaw E. Prilepsky, and Sergei K. Turitsyn, "Artificial neural networks for photonic applications—from algorithms to implementation: tutorial," Adv. Opt. Photon. 15, 739-834 (2023)

\bibitem{nn-optics}
Hadad, B. ; Froim, S.; Yosef, Erez ; Giryes, Raja ;Bahabad, Alon. Deep learning in optics - a tutorial. Journal of Optics. 25.  (2023).

\bibitem{outro1}
Zhang, D.; Tan, Z. A Review of Optical Neural Networks. Appl. Sci. 2022, 12, 5338.

\bibitem{outros2}
Zhu, H.; Lin, H.; Wu, S.; Luo, W.; Zhang, H.; Zhan, Y.; Wang, X.; Liu, A.; Kwek, L.C. Quantum Computing and Machine Learning on an Integrated Photonics Platform. Information 2024, 15, 95.

\bibitem{outros3}
Palmieri, A.M., Kovlakov, E., Bianchi, F. et al. Experimental neural network enhanced quantum tomography. npj Quantum Inf 6, 20 (2020).

\bibitem{outros4}
Quek, Y., Fort, S.; Ng, H.K. Adaptive quantum state tomography with neural networks. npj Quantum Inf 7, 105 (2021).

\bibitem{Pinn-hibrido0}
K.Zubov,Z.McCarthy, Y. Ma,F. Calisto, V. Pagliarino, S. Azeglio, L. Bottero, E. Luj´ an, V. Sulzer, A. Bharambe et al., “Neuralpde: Automating physics-informed neural networks (pinns) with error approximations,” arXiv preprint arXiv:2107.09443, (2021).

\bibitem{Pinn-hibrido1}
 J. Blechschmidt and O. G. Ernst, “Three ways to solve partial  differential equations with neural networks—a review,” GAMM-Mitteilungen, vol. 44, no. 2, p. e202100006,( 2021).

\bibitem{Pinn-hibrido2}
 J. Willard, X. Jia, S. Xu, M. Steinbach, and V. Kumar, “Integrating physics-based modeling with machine learning: A survey,” arXiv preprint arXiv:2003.04919, vol. 1, no. 1, pp. 1–34, 2020.

\bibitem{MaxTegmark2020}
 Silviu-Marian Udrescu, Max Tegmark ,AI Feynman: A physics-inspired method for symbolic regression.Sci. Adv.6,eaay2631(2020)

\bibitem{Baydin-automaticdiff}
Baydin, Atilim Gunes, et al. "Automatic differentiation in machine learning: a survey." Journal of machine learning research 18.153 (2018): 1-43.


\bibitem{Fouriernn}
Li, Z., Kovachki, N.B., Azizzadenesheli, K., Liu, B., Bhattacharya, K., Stuart, A.M., and Anandkumar, A. (2020). Fourier Neural Operator for Parametric Partial Differential Equations. ArXiv, abs/2010.08895.

\bibitem{OML}
Jiao, Shuming, et al. "Optical machine learning with incoherent light and a single-pixel detector." Optics letters 44.21 (2019): 5186-5189.

\bibitem{OML2}
Ksenia Yadav, Serge Bidnyk, and Ashok Balakrishnan, "Artificial intelligence and machine learning in optics: tutorial," J. Opt. Soc. Am. B 41, 1739-1753 (2024)

\bibitem{NeuralPDE}
Zubov, Kirill, et al. "Neuralpde: Automating physics-informed neural networks (pinns) with error approximations." arXiv preprint arXiv:2107.09443 (2021).

\bibitem{IAfeymann}
Silviu-Marian Udrescu, Max Tegmark, AI Feynman: A physics-inspired method for symbolic regression.Sci. Adv.6,eaay2631(2020).DOI:10.1126/sciadv.aay2631

\bibitem{PhyCV}
Zhou, Yiming, et al. "PhyCV: the first physics-inspired computer vision library." arXiv preprint arXiv:2301.12531 (2023).

\bibitem{PICV}
Banerjee, Chayan, et al. "Physics-informed computer vision: A review and perspectives." ACM Computing Surveys (2024).

\bibitem{Ray2023ChatGPTAC}
Ray, Partha Pratim. “ChatGPT: A comprehensive review on background, applications, key challenges, bias, ethics, limitations and future scope.” Internet of Things and Cyber-Physical Systems (2023)


\bibitem{imagem-nn}
N. Jmour, S. Zayen and A. Abdelkrim, "Convolutional neural networks for image classification," 2018 International Conference on Advanced Systems and Electric Technologies (IC ASET), Hammamet, Tunisia, pp. 397-402,(2018).

\bibitem{sound-nn}
Lin, Y.-K.; Su, M.-C.; Hsieh, Y.-Z. The Application and Improvement of Deep Neural Networks in Environmental Sound Recognition. Appl. Sci., 10, 5965, (2020)

\bibitem{text-nn}
YiTao Zhou, "Natural Language Processing with Improved Deep Learning Neural Networks", Scientific Programming, 8 pages, (2022).


\bibitem{scikit-learn_iris}
Fisher,R. A.. (1988). Iris. UCI Machine Learning Repository. https://doi.org/10.24432/C56C76.

\bibitem{scikit}
Pedregosa, Fabian, et al. "Scikit-learn: Machine learning in Python." the Journal of machine Learning research 12: 2825-2830, (2011)

\bibitem{glorot}
Glorot, Xavier, and Yoshua Bengio. "Understanding the difficulty of training deep feedforward neural networks." Proceedings of the thirteenth international conference on artificial intelligence and statistics. JMLR Workshop and Conference Proceedings, (2010).

\bibitem{activation1}
Szandała, Tomasz. "Review and comparison of commonly used activation functions for deep neural networks." Bio-inspired neurocomputing: 203-224, (2021).

\bibitem{activation2}
Sitzmann, Vincent, et al. "Implicit neural representations with periodic activation functions." Advances in neural information processing systems 33: 7462-7473. (2020)

\bibitem{dawid_modern_2022}
Dawid, A. et al. Modern applications of machine learning in quantum sciences. 	arXiv:2204.04198 (2022).



\bibitem{opt_genetical}
Al Tobi, Maamer Ali Saud, et al. "A review on applications of genetic algorithm for artificial neural network." International Journal of Advance Computational Engineering and Networking 4.9 (2016): 50-54.

\bibitem{simu_anneling}
Kuo, Chun Lin, Ercan Engin Kuruoglu, and Wai Kin Victor Chan. "Neural network structure optimization by simulated annealing." Entropy 24.3 (2022): 348.


\bibitem{otimization}
del Rosario, Mason. "An Overview of Stochastic Gradient Descent in Machine Learning." (2019).

\bibitem{otimization2}
Abdulkadirov, Ruslan, Pavel Lyakhov, and Nikolay Nagornov. "Survey of optimization algorithms in modern neural networks." Mathematics 11.11 (2023): 2466.

\bibitem{DissCafe}
Café de Miranda, G. Pulse modulation for enhanced control in NMR-based Quantum Devices. Universidade Federal do ABC, Janeiro 2024

\bibitem{Liashchynskyi}
Liashchynskyi, Petro, and Pavlo Liashchynskyi. "Grid search, random search, genetic algorithm: a big comparison for NAS." arXiv preprint arXiv:1912.06059 (2019).

\bibitem{Gower}
Gower, Robert, and Telecom Paris. "An overview of stochastic gradient-based methods." (2019).

\bibitem{adamopt}
Diederik P. Kingma and Jimmy Ba. Adam: A Method for Stochastic Optimization. arXiv:1412.6980, 2014.

\bibitem{Github}
Café de Miranda, G., Lima, G. G. Notebooks acompanhando o artigo. Repositório GitHub,\url{https://github.com/Coffee4MePlz/Notebooks_NN_Physics}.

\bibitem{MediumSafrin}
Safrin. Introduction to Neural Network. Medium, \url{https://medium.com/@safrin1128/introduction-to-neural-network-9fdff448f43f}. Acessado em: 15 de fevereiro de 2025.

\bibitem{Rumelhart}
D. E. Rumelhart, G. E. Hinton, and R. J. Williams, “Learning internal representations by error propagation,” California UnivSan Diego La Jolla Inst for Cognitive Science, Tech. Rep., (1985).

\bibitem{LeCun}
LeCun, Y., Bengio, Y. Hinton, G. Deep learning. Nature 521, 436–444 (2015).


\bibitem{Hochreiter}
 S. Hochreiter and J. Schmidhuber, “Long short-term memory,” Neural computation, vol. 9, no. 8, pp. 1735–1780, (1997).

\bibitem{transforme}
Lin, Tianyang, et al. "A survey of transformers." AI open 3: 111-132 (2022).

\bibitem{Michelucci}
Michelucci, Umberto. "An introduction to autoencoders." arXiv preprint arXiv:2201.03898 (2022).

\bibitem{Alom}
Alom MZ, Taha TM, Yakopcic C, et al, A state-of-the-art survey on deep learning theory and architectures. Electronics 8(3), (2019).

\bibitem{Herberg}
Herberg, Evelyn. "Neural Network Architectures." arXiv preprint arXiv:2304.05133 (2023).

\bibitem{hopfield}
Ramsauer, Hubert, et al. "Hopfield networks is all you need." arXiv preprint arXiv:2008.02217 (2020).

\bibitem{KAN2024}
Liu, Z. et al. KAN: Kolmogorov-Arnold Networks. Preprint at http://arxiv.org/abs/2404.19756 (2024).

\bibitem{pytoch}
Paszke, Adam, et al. "Pytorch: An imperative style, high-performance deep learning library." Advances in neural information processing systems 32 (2019).

\bibitem{tensorflow}
Martín Abadi, Ashish Agarwal, Paul Barham, et al. TensorFlow: Large-scale machine learning on heterogeneous systems, Software available from tensorflow.org. (2015)

\bibitem{jax}
Bradbury, James, et al. "JAX: composable transformations of Python+ NumPy programs." (2018).

\bibitem{autodiff}
Baydin, Atilim Gunes, et al. "Automatic differentiation in machine learning: a survey." Journal of machine learning research 18.153 (2018): 1-43.

\bibitem{MediumKhangPham}
Pham, K. Artigos no Medium. Medium, \url{https://medium.com/@khang.pham.exxact}. Acessado em: 15 de fevereiro de 2025.


\bibitem{NIPS2012_c399862d}
Alex Krizhevsky, Ilya Sutskever, Geoffrey E Hinton. "ImageNet Classification with Deep Convolutional Neural Networks". In: \textit{Advances in Neural Information Processing Systems}, vol. 25, ed. by F. Pereira et al., Curran Associates, Inc., 2012.

\bibitem{bojarski2017}
Mariusz Bojarski, Anna Choromanska, Krzysztof Choromanski, Bernhard Firner, Larry Jackel, Urs Muller, Karol Zieba. "VisualBackProp: efficient visualization of CNNs". (2017). arXiv preprint arXiv:1611.05418.


\bibitem{Lindeberg1998}
Lindeberg, T. Feature Detection with Automatic Scale Selection. Int. J. Comput. Vis. 30, 79–116 (1998).

\bibitem{EdgeDetec2008}
Park, J.M. and Murphey, Y.L. (2008). Edge Detection in Grayscale, Color, and Range Images. In Wiley Encyclopedia of Computer Science and Engineering, B.W. Wah (Ed.).

\bibitem{LindebergScaleSpace1993}
Lindeberg, T. Detecting salient blob-like image structures and their scales with a scale-space primal sketch: A method for focus-of-attention. Int. J. Comput. Vis. 11, 283–318 (1993).

 \bibitem{FukushimaCNN1980}
 Fukushima, K. Neocognitron: A self-organizing neural network model for a mechanism of pattern recognition unaffected by shift in position. Biol. Cybernetics 36, 193–202 (1980).

\bibitem{GradAmplif2020}
S. Basodi, C. Ji, H. Zhang and Y. Pan. Gradient amplification: An efficient way to train deep neural networks. Big Data Mining and Analytics, vol. 3, no. 3, pp. 196-207, Sept. 2020, doi: 10.26599/BDMA.2020.9020004.

\bibitem{Bashir}
Bashir, D., Montañez, G.D., Sehra, S., Segura, P.S., Lauw, J. (2020). An Information-Theoretic Perspective on Overfitting and Underfitting. In: Gallagher, M., Moustafa, N., Lakshika, E. (eds) AI 2020: Advances in Artificial Intelligence. AI 2020. Lecture Notes in Computer Science(), vol 12576. Springer, Cham


\bibitem{Huang-mismatch}
Huang, Chen, et al. "Addressing the loss-metric mismatch with adaptive loss alignment." International conference on machine learning. PMLR, 2019.

\bibitem{Ainsworth}
Ainsworth, Mark, and Yeonjong Shin. "Plateau phenomenon in gradient descent training of RELU networks: Explanation, quantification, and avoidance." SIAM Journal on Scientific Computing 43.5 (2021): A3438-A3468.

\bibitem{Hanin}
Hanin, Boris. "Which neural net architectures give rise to exploding and vanishing gradients?." Advances in neural information processing systems 31 (2018).

\bibitem{Lusch_KoopmanAutoencoders2018}
 Lusch, B., Kutz, J. N. and Brunton, S. L. Deep learning for universal linear embeddings of nonlinear dynamics. Nat Commun 9, 4950 (2018).

\bibitem{Gamelin2001}
Gamelin, T. W. Complex Analysis. Springer New York, New York, NY (2001). doi:10.1007/978-0-387-21607-2.

\bibitem{Gheller_2021}
C. Gheller, F. Vazza. Convolutional deep denoising autoencoders for radio astronomical images. \textit{Monthly Notices of the Royal Astronomical Society}, vol. 509, no. 1, pp. 990–1009, Oct. 2021. ISSN: 1365-2966. Available at: \url{http://dx.doi.org/10.1093/mnras/stab3044}.

\bibitem{Vincent2008}
Vincent, P., Larochelle, H., Bengio, Y. and Manzagol, P.-A. Extracting and composing robust features with denoising autoencoders. in Proceedings of the 25th international conference on Machine learning - ICML ’08 1096–1103 ACM Press (2008). doi:10.1145/1390156.1390294.

\bibitem{Geller2022}
C Gheller, F Vazza, Convolutional deep denoising autoencoders for radio astronomical images, Monthly Notices of the Royal Astronomical Society, Volume 509, Issue 1, January 2022, Pages 990–1009, https://doi.org/10.1093/mnras/stab3044

\bibitem{MSchuld2021}
 Schuld, M. and Petruccione, F. Machine Learning with Quantum Computers (2021).

\bibitem{GithubSINDy}
Sillano, Pietro. SINDy Pendulum. Github repository \url{https://github.com/pietro-sillano/SindyPendulum}

\end{thebibliography}
\end{document}